\documentclass[journal]{IEEEtran}
\IEEEoverridecommandlockouts                              

\usepackage{graphicx} 
\usepackage{amsmath} 
\usepackage[font=small]{caption}
\usepackage{subcaption}
\usepackage{mathrsfs}
\usepackage{bm}
\usepackage{amsfonts} 
\usepackage[usenames, dvipsnames]{color}
\usepackage{cite}
\usepackage{url}

\begin{document}

\title{Plugo: a VLC Systematic Perspective of Large-scale Indoor Localization}

\author{Qing Liang$^{1,2}$ and Ming Liu$^{2}$
\thanks{$^{*}$This work was sponsored by the Shenzhen Science, Technology and Innovation Comission (SZSTI) JCYJ20160428154842603 and JCYJ20160401100022706; Research Grant Council of Hong Kong SAR Government, China, under project No. 16212815, 21202816 and National Natural Science Foundation of China No. 6140021318; partially supported by the HKUST Project IGN16EG12, all awarded to Prof. Ming Liu.}
\thanks{$^{1}$Qing Liang is with Department of MBE, City University of Hong Kong.
		{\tt\small qing.liang@my.cityu.edu.hk}}%
\thanks{$^{2}$Qing Liang and Ming Liu are with Department of ECE, Hong Kong University of Science and Technology. 
		{\tt\small eelium@ust.hk}}%
}

\maketitle

\begin{abstract}
Indoor localization based on Visible Light Communication (VLC) has been in favor with both the academia and industry for years. In this paper, we present a prototyping photodiode-based VLC system towards large-scale localization. Specially, we give in-depth analysis of the design constraints and considerations for large-scale indoor localization research. After that we identify the key enablers for such systems: 1) distributed architecture, 2) one-way communication, and 3) random multiple access. Accordingly, we propose \textit{Plugo} --- a photodiode-based VLC system conforming to the aforementioned criteria. We present a compact design of the VLC-compatible LED bulbs featuring plug-and-go use-cases. The basic framed slotted Additive Links On-line Hawaii Area (ALOHA) is exploited to achieve random multiple access over the shared optical medium. We show its effectiveness in beacon broadcasting by experiments, and further demonstrate its scalability to large-scale scenarios through simulations. Finally, preliminary localization experiments are conducted using fingerprinting-based methods in a customized testbed, achieving an average accuracy of $0.14m$ along with a 90-percentile accuracy of $0.33m$.
\end{abstract}


\section{Introduction}
Location awareness is critical to many indoor applications \cite{kuo2014luxapose,armstrong2013visible,lumicast}, e.g., way-finding in a metro station, industrial unmanned ground vehicle (UGV) navigation in a warehouse, and location-based services (LBS) in retailing. The global indoor location market is projected to grow to USD 23.13 billion by 2021 \cite{marketreport}. While GPS has effectively solved the localization problems outdoors, a seamless indoor localization solution remains challenging, which is expected to be accurate, responsive, lightweight, scalable, robust, low-cost and ubiquitous\cite{lumicast,liu2007survey}. Previously, various RF-based localization methods \cite{huang2011efficient,sun2014wifi} were proposed and normally achieved accuracies around several meters owing to the multipath fading effect \cite{liu2007survey}. In the robotics community, approaches like Simultaneously Localization and Mapping (SLAM) \cite{Cadena2016} have been pervasively studied, as localization is fundamental to many robotic problems such as navigation \cite{zhong2016efficient} and path planning \cite{yang2004neural}. Centimeter-level accuracies are achievable through a laser-SLAM technique yet at the expense of high-end laser sensors. The desired indoor localization solution for widespread adoptions should seek a reasonable balance between the system performance and cost. VLC-based localization is seen as a very promising technology to fill this gap \cite{armstrong2013visible,do2016depth}. 

Unlike RF-based methods that require dedicated wireless access points (e.g., WiFi ad-hocs and Bluetooth beacons) to communication or localization, VLC-based methods leverage the existing LED lighting infrastructure which ubiquitously serves its primary function of illumination in many indoor scenarios. Moreover, a better localization accuracy could be expected as VLC signals are highly directional and immune to multipath fading effects that used to be suffered by RF signals. The dense deployment of lighting fixtures also contributes to better localization performance. In fact, we witness that some giant corporations have devoted to the R\&D on this technique, e.g., Qualcomm, Phillips, GE, and Acuity. To grant reliable localization performance, the end user requires a set of simultaneous observations of multiple light beacons. A fundamental problem is how to achieve reliable beacon broadcasting over the shared optical medium on top of commercial LED lights. 

The literature on VLC-based localization can be generally divided into camera- and photodiode-based directions with respect to the adopted VLC receivers \cite{do2016depth}. Camera-based methods of the state-of-the-art (e.g., Lumicast \cite{lumicast}) could offer high-accuracy 3D position and heading information on a commercial smartphone with an inbuilt front-facing camera using angle-of-arrival (AOA) localization algorithms. In their context, beacon broadcasting is naturally enabled when all the lights keep transmitting. This is thanks to the spatial discrimination of cameras as they are inherently AOA sensors. Nonetheless, they suffer from some intrinsic shortcomings, e.g., power consuming \cite{yang2015wearables}, computational demanding \cite{kuo2014luxapose}, relying strictly on line-of-sight (LOS) views, and requiring location registration for all the light beacons. 

On the contrary, photodiode-based methods enjoy the following benefits in nature: 1) energy-efficient thus not draining the battery hard; 2) lightweight since the signal processing is much simpler than that of camera-based methods; 3) LOS-view independent and free of location registration when using model-free localization methods (e.g., fingerprinting). To differentiate spatially clustered lights within the sensor FOV, however, extra multiple access schemes should be exploited due to the lack of spatial discrimination of photodiodes. Previously, a variety of photodiode-based localization systems were proposed through simulation and experiments \cite{li2014epsilon,zhang2014asynchronous,qiu2016let}, and different kinds of multiple access schemes were introduced \cite{rajagopal2012ieee,zhang2015asynchronous,chen2015framework}. Most of these studies evaluated the system performance in a room-size or table-size testbed, while only a few considered the scalability in large-scale scenarios. 

In this paper, we focus on the fundamental beacon broadcasting problems of a photodiode-based localization system for large-scale environments. Stimulated by Epsilon \cite{li2014epsilon} especially the random multiple access scheme, we especially take into consideration the design constraints arising from commercial LED lights with emphasis on the requirements of the large-scale indoor localization. We propose \textbf{Plugo} (named after ``Plug and Go"), a dedicated VLC system which is capable of providing reliable beacon broadcasting over a shared optical medium from multiple LED bulbs to a single photodiode receiver. It deviates from the general VLC systems \cite{karunatilaka2015led} developed in the wireless communication community that pursues high data throughput along with bidirectional communication. The key differentiation points of Plugo are threefold: 1) distributed architecture, 2) one-way communication, and 3) random multiple access.  Compared with our previous work \cite{qiu2016let,qiu2015visible}, Plugo moves a small step further towards the expected localization technology.

Specifically, we stress our novel contributions as follows:
\begin{enumerate}
\item Identifying the key enablers to a photodiode-based VLC system towards large-scale localization through in-depth analysis of the design constraints and considerations;
\item Design, implementation, and evaluation of such a VLC system that is easy to use in a plug-and-go fashion;
\item Implementation of a basic framed slotted ALOHA\footnote{ALOHA stands for Additive Links On-line Hawaii Area. It was originally proposed in 1970s for wireless medium access control.} (BFSA) random multiple access scheme with practical issues taken into account. This is the first experimental demonstration in this community to our knowledge. 
\end{enumerate}
 
The remainder of this paper is organized as follows. Section II discusses from scratch the design constraints and considerations; The detailed system design and implementation are described in Section III and Section IV respectively; Section V demonstrates the evaluation results; Finally, we conclude this paper in Section VI.   

\section{Design Constraints and Considerations}

\subsection{Commercial LED Lights}
LED lights comprise a number of high-powered white LEDs, which can be switched on and off rapidly to convey information via VLC at a high frequency imperceivable to human eyes \cite{karunatilaka2015led}. The basic principle of VLC is intensity modulation and direct detection (IM/DD). This is because visible light emitted by white LEDs is inherently noncoherent light with a broad spectrum. It is unlikely to modulate either the frequency or phase of visible light itself. The most widely adopted white LED in luminaries is made of a blue LED with a yellow phosphor coating. It is much cheaper and more energy-efficient than the RGB-type LEDs. The typical modulation bandwidth is around 2 MHz which is more than enough for localization purpose \cite{armstrong2013visible}. However, there are still some inherent constraints. For example, the color shift keying (CSK) modulation and wavelength division multiple access (WDMA) schemes only work with RGB-type LEDs and thus should be avoided in a system for widespread adoptions. To fulfill the primary illumination function of LED lights, we have to tackle the potential flicker and dimming issues.

\subsection{Large-scale Localization}
The long-term goal is to build a commercially viable localization system for widespread adoptions in large-scale indoor scenarios with minimized requirements of installation and configuration. We identify the key technical criteria to enable this on the basis of \cite{lumicast} and \cite{liu2007survey} in the following:

\textbf{Accurate}. It concerns both the accuracy and precision. Normally, the accuracy is measured by the mean distance error while the precision described by the empirical cumulative distribution function (CDF). This is a straightforward requirement of many indoor applications like LBS.

\textbf{Responsive}. It means a short localization latency and a high update rate. This is essential for real-time use cases such as UGV navigation and pedestrian tracking.  

\textbf{Lightweight}. It refers to the minimized computational overhead and system complexity along with power consumption. In this way, compelling indoor location services are likely to be affordable by resource-constrained mobile devices without draining heavily either the computation resources or battery.

\textbf{Scalable}. The meaning is twofold: 1) scaling to the coverage area and 2) scaling to the user density. In other words, the localization performance should not degrade significantly as the localization scope increases or more users are involved. This is critical to large-scale localization. 

\textbf{Robust}. The system could function well in some harsh scenarios, e.g., with some beacons broken abruptly or coming into operation for the first time.

\textbf{Low-cost}. The cost concerns money, time, labor, etc. It is challenging to seek an elegant balance between the performance and the cost. 

\textbf{Ubiquitous}. The location service can be generally enabled in most scenarios once required. It is the ultimate goal to provide a ubiquitous indoor location service which is comparable to what GPS has done outdoors.

Beacon broadcasting on top of LED lights founds the basis of VLC-based localization. To build a photodiode-based VLC system conforming to the above criteria for large-scale localization, we give in-depth analysis of the design considerations mainly from three aspects: network architecture, communication links, and multiple access schemes. 

\subsection{Centralized vs. Distributed Architecture}
For a centralized architecture, beacons are scheduled by a superior coordinator through a wired or wireless backbone network. To grant a strict time synchronization, localization methods based on time-of-arrival (TOA) and time-difference-of-arrival (TDOA) normally require a wired backbone connection among the transmitters \cite{do2016depth}. Besides, methods that exploit the time division multiple access (TDMA) also rely on a centralized architecture \cite{yasir2014indoor}. However, the backbone network increases the total hardware complexity along with the installation cost. Worse still, a centralized system is prone to collapse once the coordinator is broken. On the contrary, the distributed architecture is free of either a coordinator or a backbone connection. It is likely to provide improved robustness during the operation, reduced hardware complexity and cost during the installation. Many localization systems were proposed in the literature with a distributed architecture \cite{kuo2014luxapose,li2014epsilon,zhang2014asynchronous,qiu2016let}. 

\subsection{Two-way vs. One-way Communication}
From now on, we consider a distributed architecture of the light beacons. As for a broadcasting network, communication between the beacon and mobile user can be either bidirectional or unidirectional. Contention based multiple access schemes can be exploited on top of two-way communication, e.g., Carrier Sense Multiple Access with Collision Avoidance (CSMA/CA). Time synchronization is also feasible so as to favor a time-of-flight (ToF) localization method \cite{liu2007survey}. Several networked VLC systems were proposed in the communication community that exploited a CSMA/CA-based medium access control (MAC) \cite{wang2015open,klaver2015shine}. From the localization standpoint, however, systems relying on two-way communication between the infrastructure and the end user do not scale well as the beacons and users increase. This is because the frequent handshake connections in two-way communication burden the total traffic thus involving more interference to other nodes working concurrently. Besides, the hardware complexity along with cost will increase as two-way communication requires VLC transceivers on both the beacon and mobile user sides. To enable large-scale localization, we prefer one-way communication because of the better scalability and reduced hardware complexity as well as cost. In fact, some camera-based commercial localization systems offered by Qualcomm \cite{lumicast} and Phillips \cite{philips} support one-way communication only.  

\subsection{Fixed vs. Random Multiple Access}
We now focus on a fully distributed VLC system with one-way communication. Due to the absence of either a backbone network or bidirectional communication, it's no longer possible to perform any synchronization either between each two beacons or between the beacon and the mobile user. To tackle this problem, some asynchronous multiple access schemes have been proposed, e.g., frequency division multiple access (FDMA) \cite{kim2013indoor} and code division multiple access (CDMA) \cite{liu2014towards}. Normally, a fixed frequency carrier or pseudo-noise sequence is allocated to each light beacon in advance. When it comes to a large-scale environment with thousands of lights, however, the number of frequency carriers or pseudo-noise sequences available are limited to use. Authors in \cite{kim2013indoor} proposed an RF-carrier allocation scheme to mitigate the inter-cell interference by reusing a limited number of carriers in non-adjacent cells. In their context, one had to guarantee that none of any adjacent lights shared the same frequency carrier. It imposed a stringent constraint on the practical deployment, especially in a large-scale environment. Indeed, this is a common problem of many other fixed multiple access schemes (e.g., CDMA), but it has long been overlooked by many researchers. 
 
A radical idea is to dynamically assign a limited number of communication resources (e.g., time slots, frequency carriers, and optical CDMA codes) in a random fashion to all the beacons involved --- random multiple access \cite{jplrandom}. In this context, every beacon in the VLC system competes with one another equally for communication resources. As a result, the random access scheme is free of the headache suffered by its counterpart. Collisions occur when two different beacons compete for the same piece of resources (e.g., time slots). But this problem can be easily worked out through multiple observations. To the best of our knowledge, Epsilon \cite{li2014epsilon} was the first experimental system in this community that involved random multiple access --- channel hopping. A BFSA-based random access scheme was first introduced by Zhang et al \cite{zhang2014asynchronous} and evaluated through simulation. In this paper, we will demonstrate the implementation of this scheme by taking into account practical issues.

\section{The Plugo System}
\begin{figure}[thpb]
	\centering  
    \includegraphics[width=0.95\columnwidth]{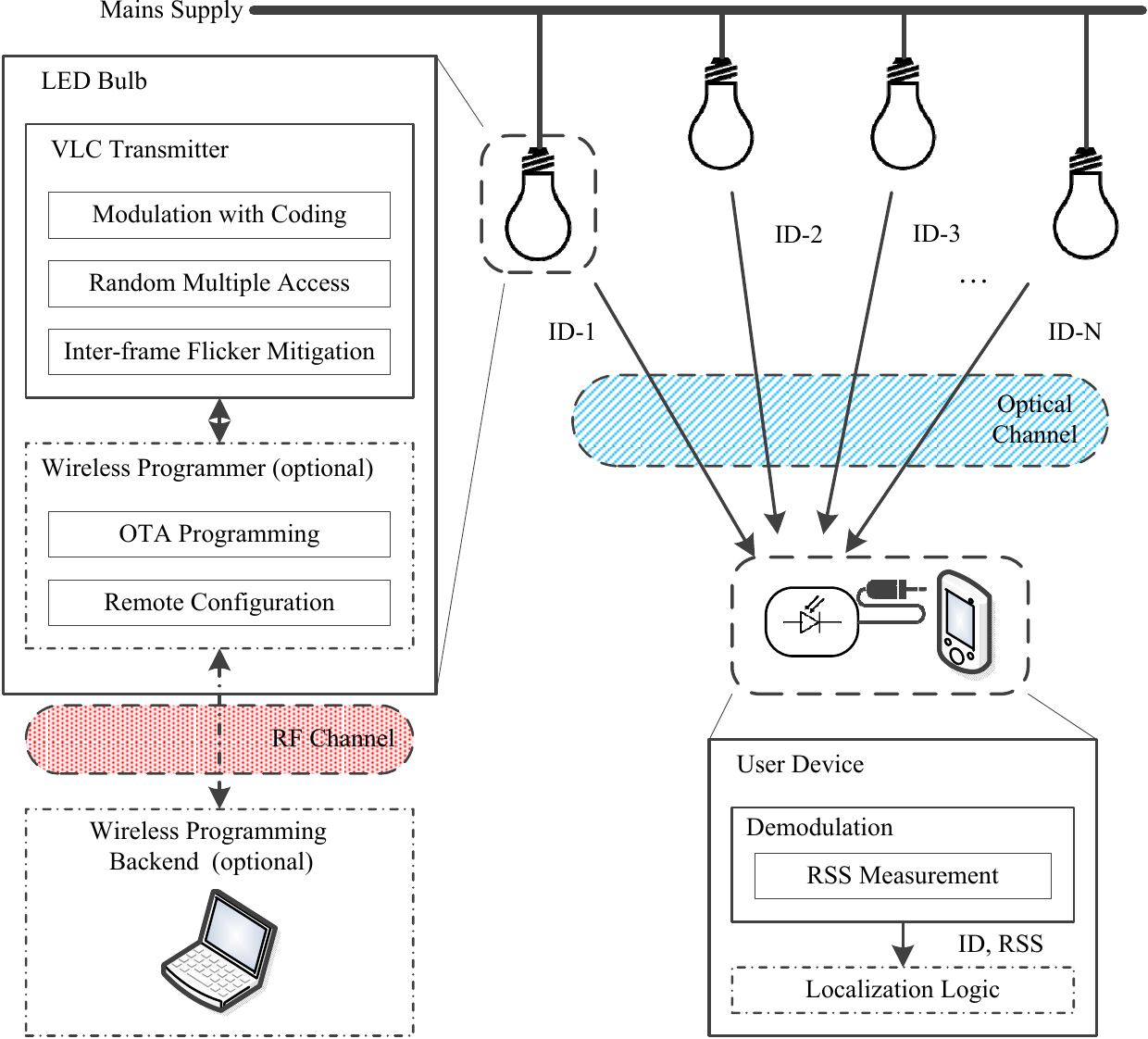}      
    \caption{The Plugo System Architecture.
    }
    \label{fig:system}
\end{figure}

The Plugo system, as shown in Figure \ref{fig:system}, comprises a set of VLC-compatible LED bulbs as positioning beacons and a photodiode receiver which is attached to the user device (e.g., smartphones, tablets) via an audio jack. The system architecture is distributed without any backbone connections among these bulbs. Each bulb broadcasts its unique beacon identity to the receiver by one-way communication over a shared optical channel. The user device equipped with a receiver takes continuous observations of multiple light beacons overhead retrieving each beacon identity along with the corresponding received signal strength (RSS). VLC-based localization is thus feasible on the top of these observations.

Each bulb exploits a microcontroller to implement all the communication logics as a firmware such as signal modulation and multiple access control. The beacon message is encoded by simple on-off keying (OOK) modulation with Manchester coding. Subject to one-way communication, the bulb could never know whether the sent beacon message has been correctly interpreted or not by the receiver. We adopt a BFSA-based random multiple access scheme in Plugo to prevent persistent collisions among these ``blind" and uncoordinated light beacons. In addition, we have solved the induced inter-frame flicker problem. 

To ease the system debugging such as the firmware update and parameter configuration, we come up with a standalone wireless programming system exploiting the RF channel. It consists of a backend and a number of wireless programmers residing in LED bulbs. Over-the-air (OTA) programming and remote configuration can be thus achieved. However, it should be highlighted that the RF links along with the wireless programming backend are not involved in beacon broadcasting. They only serve for debugging purpose.

\subsection{Communication using OOK}

\subsubsection{Modulation and Coding}
Many modulation schemes have been proposed for VLC, e.g., OOK, VPPM (Variable Pulse Position Modulation), and OFDM (Orthogonal Frequency Division Multiplexing). As for beacon broadcasting, a high data throughput is not necessary. We prefer a simple modulation scheme affordable in low-cost hardware components. To be specific, we choose the OOK modulation with Manchester coding which is also adopted by the IEEE 802.15.7 PHY I layer specification. It favors DC balance and easy clock recovery. While the binary frequency shift keying (BFSK) used by Epsilon \cite{li2014epsilon} requires the FFT operation during the demodulation, OOK is even more lightweight as it can be demodulated on a low-cost microcontroller without DSP units. Constraints on the modulation frequency $f_{mod}$ come from many factors such as the response time of LEDs, the perception bandwidth of photodiodes, the sampling frequency of the Analog-to-digital converter (ADC), etc. As for a proof-of-concept system implementation, we empirically choose $f_{mod}=10 \, \mathrm{kHz}$ which is interpretable by a USB soundcard with a maximum sampling frequency of 48kHz.

\subsubsection{Protocol Definition}

\begin{figure}[thpb]
	\centering
    \includegraphics[scale=0.5]{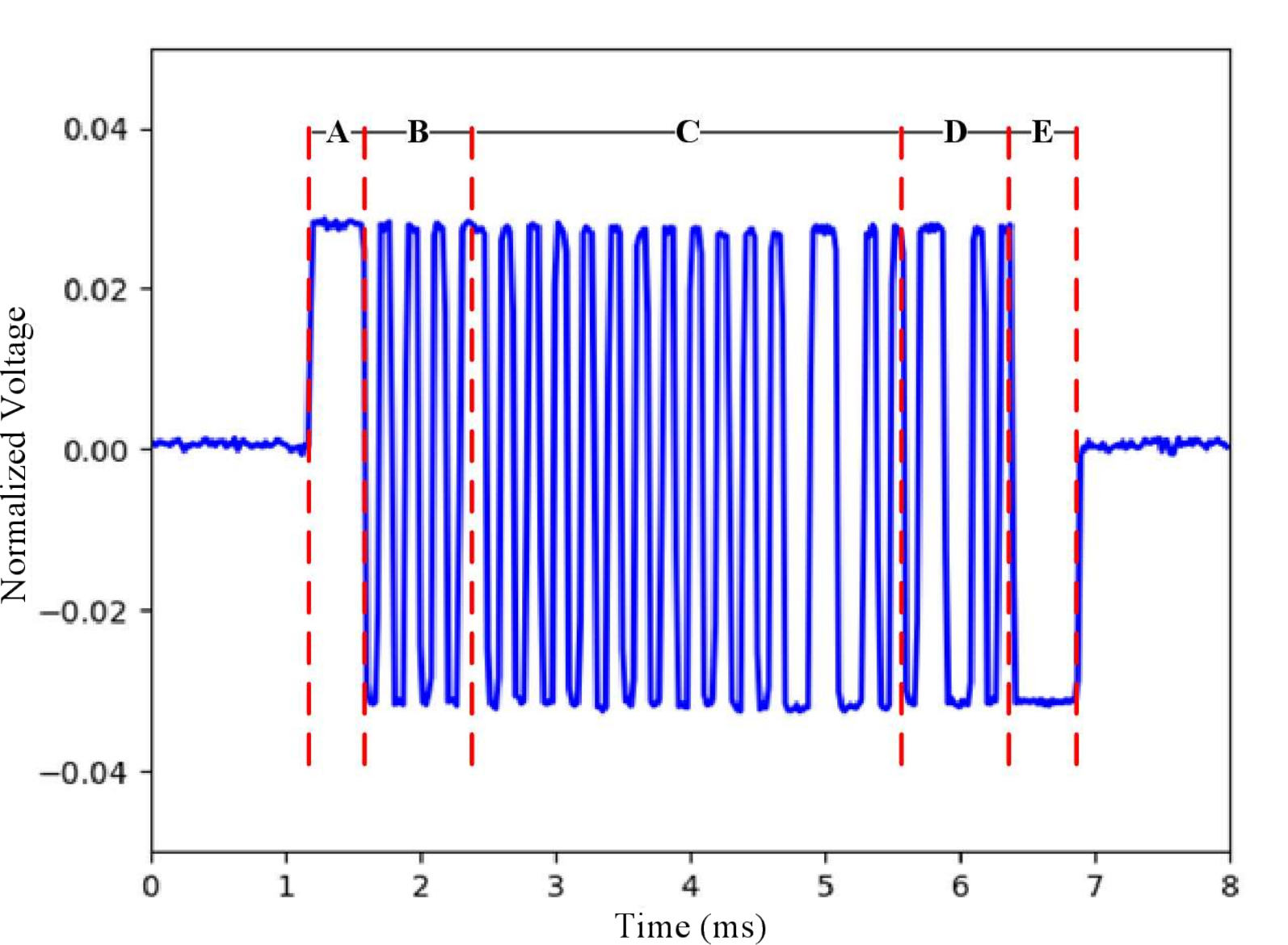}     
    \caption{A received VLC signal sample containing a full data frame. It is shown in a bipolar form as we adopt a USB sound card with AC coupling to acquire the analogue signal from the receiver front end. The signal amplitude is normalized within [-1.0, 1.0]. Notations A to E: A-SFD, B-Sync, C-Data Payload, D-Data Checksum, E-EOF.  
    }
    \label{fig:sig_frame}
\end{figure}

Due to the limited modulation bandwidth of the low-cost hardware components, we do not chase a complete communication protocol with forward error correction (FEC) channel coding (e.g., Reed-Solomon and convolutional coding) and sophisticated MAC control mechanisms. Instead, we design a simple data frame structure that conforms to the beacon broadcasting application, as shown in Table \ref{tab:data_frame}. It is composed of three sections, namely, start-of-frame (SOF), Data and end-of-frame (EOF). Figure \ref{fig:sig_frame} illustrates the frame composition using a sample of raw received VLC signals. SOF further contains a special frame delimiter (SFD) to indicate the start of a new frame and a Sync sequence for clock synchronization. The SFD here is indeed a 4-bit logic high symbol that never occurs in the normal Manchester coding data. The Sync sequence is 8-bit long with alternate high and low logic symbols that carry timing information. Similarly, the EOF contains a 4-bit logic low symbol ($\sim$SFD) to indicate the end of a frame. The data section consists of a 16-bit payload and a 4-bit checksum, which are both encoded by Manchester coding. The payload carries a unique identification code for each bulb. A 16-bit long code could easily cover a normal indoor environment with tens of thousands of lights. The checksum is generated by a simple XOR operation to verify the payload integrity. Once data corruption is detected at the receiver side, the message will be discarded. We do not perform any data retransmission under this circumstance, as the light beacon could not detect the potential transmission failure due to the lack of an uplink. Instead, the corrupted message could be recovered by subsequent observations. 

\begin{table}[thpb]
\centering
\caption{Data Frame Structure}
\label{tab:data_frame}
\begin{tabular}{|c|c|c|c|c|}
\hline
\multicolumn{2}{|c|}{\textbf{SOF}} & \multicolumn{2}{c|}{\textbf{Data}} & \textbf{EOF} \\ \hline
SFD              & Sync            & Payload         & Checksum         & $\sim$SFD    \\ \hline
4 bits           & 8 bits          & 32 bits         & 8 bits           & 4 bits       \\ \hline
\end{tabular}
\end{table}

\subsection{Random Multiple Access}
Random multiple access is the key difference between Plugo and many other systems that exploit fixed multiple access schemes such as FDMA and CDMA. To the best of our knowledge, it was only adopted by Epsilon \cite{li2014epsilon} and Zhang's work \cite{zhang2014asynchronous} in the previous literature. Epsilon proposed a frequency channel hopping scheme and showed its effectiveness through experiments. Zhang et al. introduced the BFSA scheme by simulation. It defines a transmission frame structure composed of a fixed number of time slots. Each light beacon selects randomly a time slot in the frame and transmits its beacon message within that slot. If any two adjacent lights select the same time slot or two overlapping time slots, there will be a transmission failure because of the conflicts. However, the corrupted messages could be safely recovered later.

We adopt the BFSA scheme to achieve random multiple access over the shared optical medium in a distributed VLC network that supports one-way communication. The success rate $P_{success}$ is an important metric to evaluate the system performance. It is defined in \cite{zhang2014asynchronous} as the probability of successful transmission from all transmitters inside a coverage vicinity to a given receiver. Supposing perfect synchronization among these transmitters, the theoretical value could be calculated by the following formula where $N$ is the number of time slots per frame and $n$ is the number of transmitters. 

\def \bangle{ \atopwithdelims \langle \rangle}

$$P_{success}=\frac{{{N}\bangle{n}}}{N^{n}}$$

As for an asynchronous system, starting points of the time slots from different lights are probably misaligned. The success rate will decrease when more collisions happen. $P_{success}$ increases with the number of time slots per frame $N$. Meanwhile, the communication bandwidth needed for each transmitter is also $N$ times the original. We have to make a trade-off between the success rate and communication bandwidth available on low-cost hardware components.
  
The LED bulb performs data transmission when the desired time slot comes and then goes to idle states. The Manchester coding data is DC balanced thus eliminating the intra-frame flicker of the bulb during the data transmission. However, how to mitigate the potential inter-frame flicker during the idle states remains a problem. 

\subsection{Inter-frame Flicker Mitigation}
Flicker refers to the visible fluctuation of the light brightness \cite{rajagopal2012ieee}. It further comprises intra-frame flicker and inter-frame flicker. In this context, the intra-frame flicker has been eliminated by Manchester coding. Thus we focus on the mitigation of the inter-frame flicker. As suggested by the IEEE 802.15.7 standard \cite{ieeevlc}, one can make the LED light transmit a dummy data message during the idle states to prevent flicker. The modulation frequency of the dummy message can either be in-band or out-of-band. In a particular case, the light can be driven by a suitable DC current free of modulation. The idea is straightforward --- the overall brightness of the LED bulb will keep consistent as long as the DC intensity during the idle time slots equals to the average intensity during the active time slots. The required DC current of each bulb varies. It is unlikely to tune it one by one.

As for the Plugo system, it is a bad way to modulate the dummy message by an in-band frequency because it can induce severe interference to other nodes who are broadcasting beacon messages. We prefer to modulate it by a high out-of-band frequency, e.g., 100 kHz in the current implementation, which can be removed easily by a low-pass filter on the receiver. The dummy message here is indeed repeated ``01" symbols which provide an equal average intensity to the beacon message. Ideally, we may also choose a much lower frequency like 100 Hz which has to be filtered out by a high-pass filter. In this context, however, the modulated OOK signal will be distorted severely by the high-pass filter and become difficult to recover. This is due to the significant attenuation of the low-frequency components of the modulated square waves.

\subsection{Localization using Gaussian Process Regression}
As for the localization validation of the Plugo system, we adopt a fingerprinting-based localization algorithm which uses Gaussian process regression (GPR) \cite{rasmussen2006gaussian} based on our previous work \cite{qiu2016let,qiu2015visible}. To be specific, GPR is utilized to construct an intensity distribution model for the environment according to sparsely collected fingerprint samples. After that, a Bayes filter is used to do localization with respect to the built map. 

\subsubsection{GPR-based Environment Modelling}
We consider a training set $\mathcal{D}=\{(\mathbf{x}_{1},y_{1}),(\mathbf{x}_{2},y_{2}),\dots,(\mathbf{x}_{n},y_{n})\}$, where $\mathbf{x}_{i}$ denotes the input vector of a 2D position, $y_{i}$ is the scalar observation of the received RSS for each light source. The observation is drawn from a noisy process $y_{i}=f(\mathbf{x}_{i})+\epsilon$, where $\epsilon$ is an additive Gaussian noise with zero mean and known variance $\sigma_{n}^{2}$. For notational convenience, all $\mathbf{x}_{i}$ are aggregated into a design matrix $\mathbf{X} = [\mathbf{x}_{1} \, \mathbf{x}_{2} \dots \mathbf{x}_{n}]^{\mathrm{T}} \in \mathbb{R}^{n\times2}$, and $y_{i}$ into a column vector $\mathbf{y} = [y_{1} \, y_{2} \dots y_{n}]^{\mathrm{T}} \in \mathbb{R}^{n}$. 

Gaussian processes (GPs) are exploited to predict the posterior distributions over functions $f$ from the training set $\mathcal{D}$. The covariance between two function values $f(x_{p})$ and $f(x_{q})$ depends on the input values $x_{p}$ and $x_{q}$, which could be characterized by the kernel function $k(\mathbf{x}_{p},\mathbf{x}_{q})$. We normally choose a Gaussian (a.k.a., RBF) kernel, $$k(\mathbf{x}_{p},\mathbf{x}_{q})=\sigma_{f}^{2} \mathrm{exp} \left[-\frac{(\mathbf{x}_{p}-\mathbf{x}_{q})^{2}}{2l^{2}} \right]$$ 
where $\sigma_{f}^{2}$ is the signal variance and $l$ the length scale. These parameters specify how strongly the two points are correlated. The covariance for the noisy observations is $cov(y_{p},y_{q})=k(\mathbf{x}_{p},\mathbf{x}_{q})+\sigma_{n}^{2}\delta_{pq}$. 
Here, $\sigma_{n}^{2}$ is the Gaussian noise and $\delta_{pq}$ is one if
$p = q$ and zero otherwise. As for the entire training set, $$cov(\mathbf{y})=\mathbf{K}+\sigma_{n}^{2}\mathbf{I}$$
where $\mathbf{K} = \left[k(\mathbf{x}_{p},\mathbf{x}_{q})\right] \in \mathbb{R}^{n\times n}$ is the covariance matrix of the input values. Given an arbitrary input $\mathbf{x}_{\ast}$, the posterior distribution over function values $f(\mathbf{x}_{\ast})$ is Gaussian, $$p\left(f(\mathbf{x}_{\ast}) \, | \, \mathbf{x}_{\ast},\mathbf{X},\mathbf{y}\right)=N\left( f(\mathbf{x}_{\ast});\mu_{\mathbf{x}_{\ast}},\sigma_{\mathbf{x}_{\ast}}^{2}\right)$$
where $\mu_{\mathbf{x}_{\ast}}=\mathbf{k}_{\ast}^{\mathrm{T}}(\mathbf{K}+\sigma_{n}^{2} \mathbf{I})^{-1} \mathbf{y}$ and $\sigma_{\mathbf{x}_{\ast}}^{2}=k(\mathbf{x}_{\ast},\mathbf{x}_{\ast})-\mathbf{k}_{\ast}^{\mathrm{T}} (\mathbf{K}+\sigma_{n}^{2} \mathbf{I})^{-1} \mathbf{k}_{\ast}$. Here, the vector $\mathbf{k}_{\ast} \in \mathbb{R}^{n}$ denotes the covariances between $\mathbf{x}_{\ast}$ and $n$ training inputs $\mathbf{X}$. So as to estimate the corresponding noisy observation $y_{\ast}$, one has to add the observation noise $\epsilon$, $$p\left(y_{\ast} \, | \, \mathbf{x}_{\ast},\mathbf{X},\mathbf{y}\right)=N\left(y_{\ast};\mu_{\mathbf{x}_{\ast}},\sigma_{\mathbf{x}_{\ast}}^{2}+\sigma_{n}^{2}\right)$$

The predictive distribution provides a probabilistic regression model of the received VLC signal strength with respect to locations. Accordingly, we can create a set of intensity distribution maps comprising the mean maps and variance maps for each light beacon for localization. 

\subsubsection{Bayes Filter-based Localization}
The formulation of localization using Bayesian filtering is,
\begin{align*}
&p(\mathbf{x}_{t} \,|\, \mathbf{y}_{0:t}, \mathbf{u}_{0:t}) \propto p(\mathbf{y}_{t} \,|\, \mathbf{x}_{t}) \\ &\times \sum_{\substack{\mathbf{x}_{t-1}}} p(\mathbf{x}_{t} \,|\, \mathbf{x}_{t-1},\mathbf{u}_{t-1})p(\mathbf{x}_{t-1} \,|\,\mathbf{y}_{0:t-1}, \mathbf{u}_{0:t-1})
\end{align*} where, $\mathbf{u}_{0:t}$ is the control input, $p(\mathbf{y}_{t} \,| \, \mathbf{x}_{t})$ is the observation model, $p(\mathbf{x}_{t} \,|\, \mathbf{x}_{t-1}, \mathbf{u}_{t-1})$ is the motion model, and $p(\mathbf{x}_{t-1} \,|\,\mathbf{y}_{0:t-1}, \mathbf{u}_{0:t-1})$ is the previous estimation. 

Assuming independence across observations from different lights, the observation model could be formulated as $p(\mathbf{y}_{t} \,| \, \mathbf{x}_{t}) = \prod_{l=1}^{L}p(\mathbf{y}_{t}^{l} \,| \, \mathbf{x}_{t})$. Here, $L$ is the total number of light beacons and $p(\mathbf{y}_{t}^{l} \,| \, \mathbf{x}_{t})$ is the observation model for each light generated by GPR. For now, we do not involve any motion measurement, e.g., IMU readings. The  motion model respects a zero-mean Gaussian distribution with known variance $\sigma^{2}$ describing the movement uncertainty. We initialize the location prior using an uniform distribution. The localization result could be solved by maximum-a-posterior (MAP),$$\hat{\mathbf{x}}_{t}=\underset{\mathbf{x}_{t}}{\operatorname{argmax}}\,\{p(\mathbf{x}_{t} \,|\, \mathbf{y}_{0:t}, \mathbf{u}_{0:t})\}$$

\section{Implementation Details}

\subsection{VLC-compatible LED Bulbs}
\begin{figure}[thpb]
	\centering	
    \includegraphics[width=0.7\columnwidth]{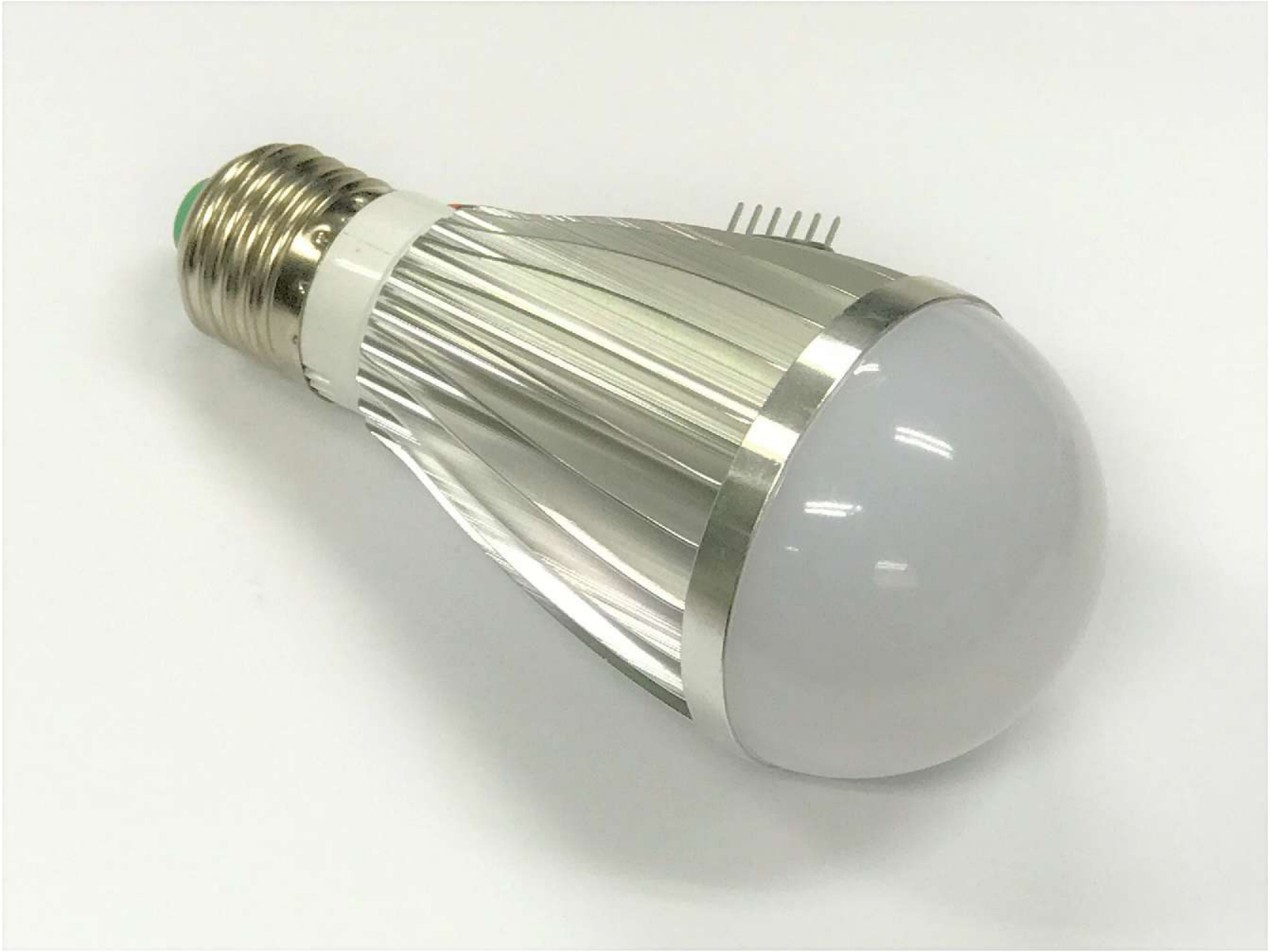}   
    \caption{The fully assembled LED bulb.}
    \label{fig:bulb_full}
\end{figure}  

\begin{figure}[thpb]
	\centering	    
    	\includegraphics[scale=1]{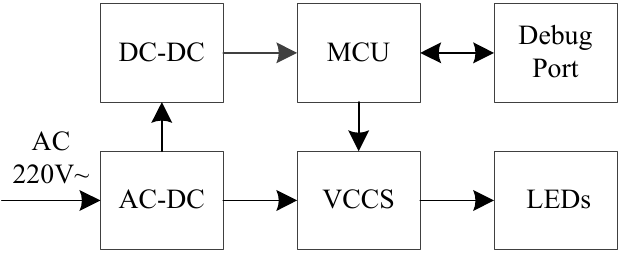}     
    \caption{Schematic diagram of the LED bulb driver circuitry
    }
    \label{fig:bulb_sch}
\end{figure}

\begin{figure}[thpb]
	\centering		
    \includegraphics[width=0.9\columnwidth]{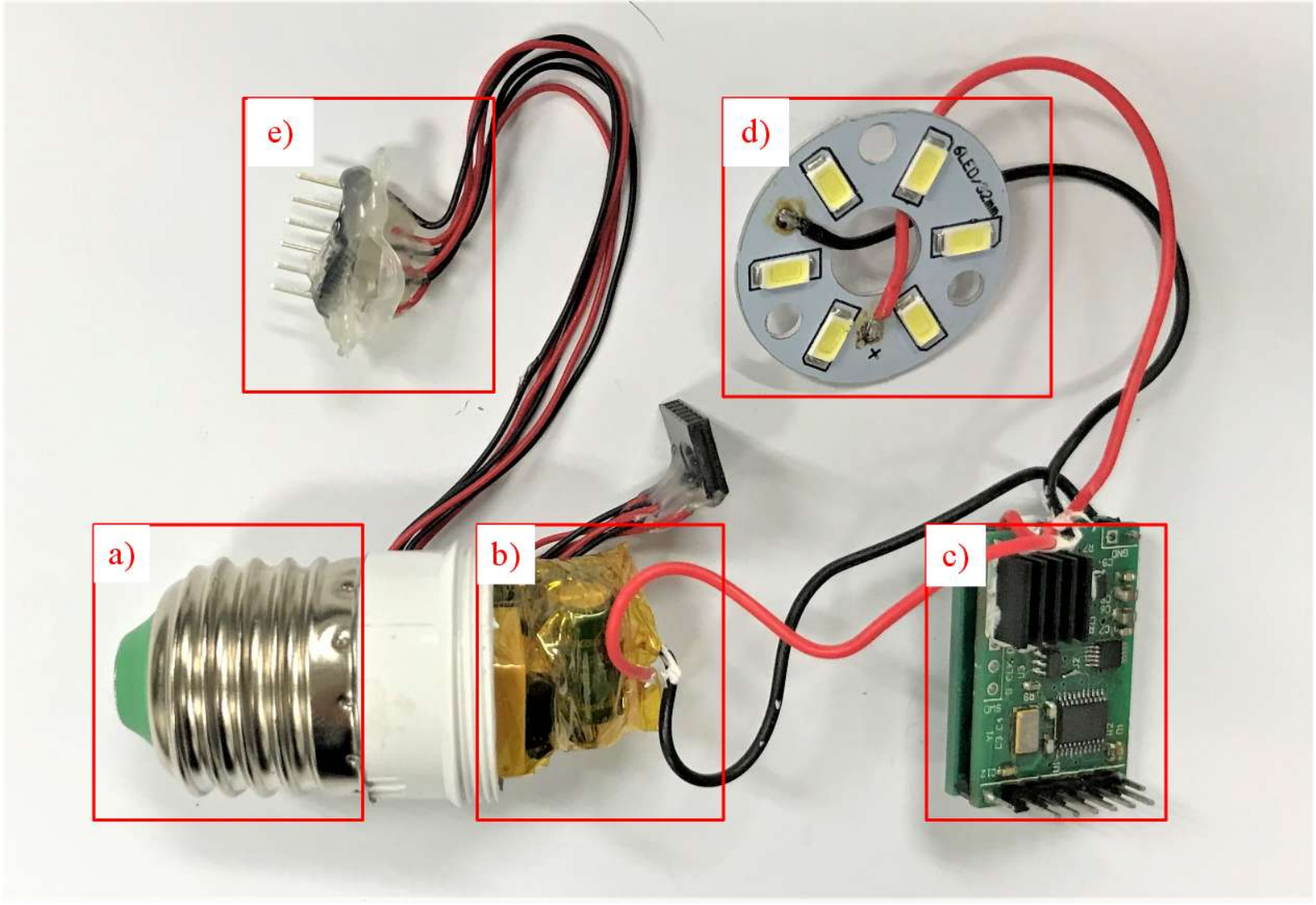}   
    \caption{The primary components of LED bulbs. a) a standard E27 screw base, b) an off-the-shelf AC-DC power supply, c) the VLC control board integrating a DC-DC converter and a microcontroller, d) a LED plate, and e) a debugging connector. The bulb case and shade are omitted for brevity.
    }
    \label{fig:bulb_comp}
\end{figure}

The VLC functionality is not yet available on off-the-shelf LED lights. As shown in Figure \ref{fig:bulb_full}, we aim at a compact design of the LED bulb which is easy to use in a plug-and-go fashion. The bulb is designed with a standard E27 screw base so that it can be easily installed to a lamp socket. The schematic of the LED driver is shown in Figure \ref{fig:bulb_sch}. It consists of an AC-DC power supply, a DC-DC buck converter, a voltage-controlled current source (VCCS), a low-cost microcontroller (MCU), a debug connector, and a LED plate. The AC-DC power module provides an output of 12 V with a maximum power of 4.5 W. We choose a 3 W LED plate, considering the power tolerance. The DC-DC converter steps down 12 V to 5 V to power other circuits. The LED current is adjusted by the VCCS under the control of the microcontroller. The signal modulation, coding, and the random multiple access control are all implemented in the microcontroller as a firmware. With the aid of a wireless programmer, the firmware can be updated over the air on demand. We build the bulbs on the top of a set of off-the-shelf LED bulb components, as shown in Figure \ref{fig:bulb_comp}. The aluminum made bulb case is good for heat dissipation. Moreover, there is enough room inside the case to hold a small driver circuitry.  

\subsection{Photodiode Receiver}

\begin{figure}[thpb]
	\centering
    \includegraphics[scale=1]{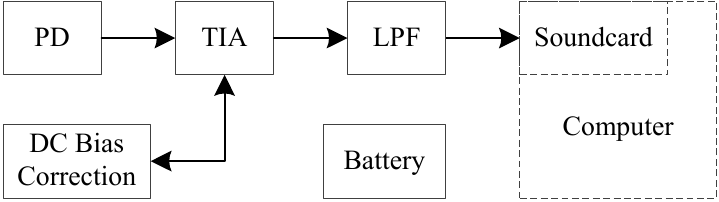}     
    \caption{Schematic diagram of the photodiode receiver circuitry
    }
    \label{fig:rec_sch}	
    \includegraphics[width=0.9\columnwidth]{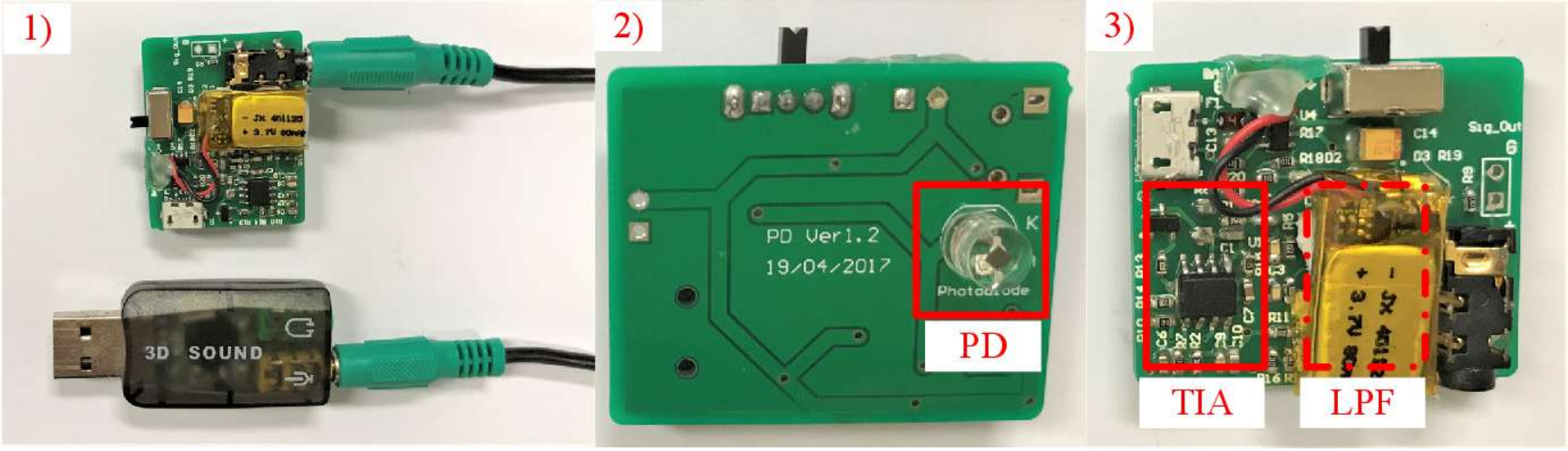}      
    \caption{(1) The receiver circuitry connected with a USB sound card via the audio jack; (2) Top view of the PCB showing the photodiode (highlighted in the red rectangle); (3) Bottom view of the PCB showing the TIA and LPF (covered by the lithium battery).
    }
    \label{fig:rec_pcb}
\end{figure}

The schematic of the designed receiver circuitry is shown in Figure \ref{fig:rec_sch}. It is composed of a PIN photodiode (PD), a trans-impedance amplifier (TIA) with DC bias correction, a low-pass filter (LPF), and a small lithium battery. We connect the receiver to a USB soundcard via an audio jack. The signal acquisition and demodulation are implemented on the computer with the \textit{python-alsaaudio} \footnote{\url{https://larsimmisch.github.io/pyalsaaudio/}} library. The lithium battery can be recharged via a micro USB connector. Figure \ref{fig:rec_pcb} shows the assembled circuitry.

The ambient interference comes from sunlight along with the fluorescent or incandescent lights. It includes a large DC bias, some strong low-frequency components (100 or 120 Hz), and high-frequency harmonics. Besides, the dummy message broadcasting in our system also introduces a significant high-frequency component. A large DC bias may cause saturation of the receiver circuitry. To circumvent this situation, we involve an error integrator to TIA so as to correct the induced DC bias. The output signal is biased to a fixed value despite of the ambient interference. We adopt a fourth-order Butterworth low-pass filter to remove the significant high-frequency interference from the dummy broadcasting. We do not take care of the low-frequency interference in the current implementation.

\subsection{Wireless Programming System}
\begin{figure}[thpb]
	\centering
    \includegraphics[width=0.9\columnwidth]{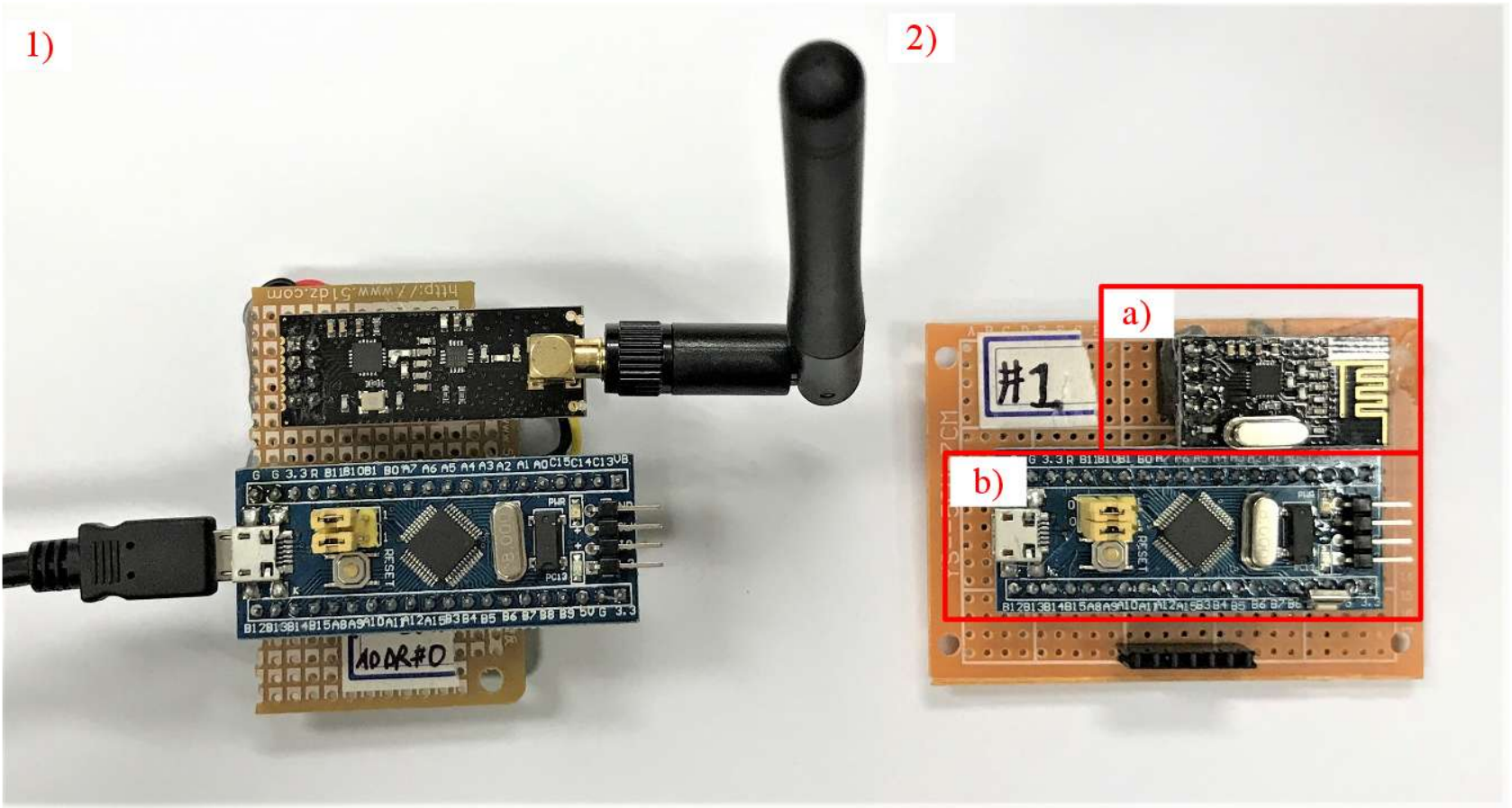}
      
    \caption{The wireless programming system prototype. (1): The master node with a high-gain antenna; (2): The slave node consisting of a RF module a) and a microcontroller boad b).
    }
    \label{fig:prog}
\end{figure}

The wireless programming system is designed to fulfill two primary functions---1) over-the-air (OTA) programming and 2) remote configuration. The motivation arises from the needs of system debugging in a large-scale environment, e.g., with tens or hundreds of VLC-compatible LED bulbs. During the debug session, we may modify the firmware of the bulbs or alter configuration parameters from time to time. It would turn to be a nightmare if we have to unscrew each bulb and flash a firmware image piece by piece. To circumvent this situation, a wireless programming system comes to play. It consists of a master node (Figure \ref{fig:prog}-1) connected with a computer as the backend and a number of slave nodes (Figure \ref{fig:prog}-2) attached to the LED bulbs. They are assembled on the basis of a shared hardware design albeit running different versions of firmware. We build several hardware prototypes using off-the-shelf components including a microcontroller board with a USB interface and a low-cost 2.4G RF communication module. To cover a broader area, the RF module in the master node bears a larger transmission power considering that the data traffic occurs mainly in the downlink.

The master node is in charge of each communication session. It could talk to any slave nodes by assigning a specific destination address so as to initiate a firmware update or parameter configuration. All the control logics are implemented as a firmware inside the microcontroller. The master node could be configured by the computer via the USB interface using a set of customized ``AT" commands. As for the OTA programming, the computer specifies the destination address along with other necessary parameters for RF communication (e.g., channel frequency) and sends the firmware image to be flashed to the master node through USB. The firmware file will be packaged and then dispatched to the designated slave node. The slave will recover the firmware image and check its integrity. If the received file is valid, the slave node will embark on the firmware downloading using in-system-programming (ISP) via UART with the target bulb. Besides, we have implemented a virtual serial communication protocol which is transparent to users on top of RF links. As a consequence, remote configuration of LED bulbs is feasible.

\section{System Evaluation}
In this section, we evaluate Plugo first with a small customized testbed through field experiments, and then simulations for large-scale scenarios. As for the field experiments, we focus on the evaluation of the basic beacon broadcasting capability of Plugo along with its localization performance in real applications. We implement a fingerprinting-based localization algorithm and present a preliminary localization result demonstrating the localization accuracy, consistency, responsiveness and robustness. To further explore the system scalability to large-scale scenarios, we conduct simulations in a floor-size environment especially evaluating the beacon broadcasting performance. As for the localization evaluation in large-scale environments, we refer the readers to our technical report\footnote{\url{https://ram-lab.com/papers/tr_plugo}} due to limited space.

\begin{figure}[thpb]
	\centering
    \includegraphics[width=0.9\columnwidth]{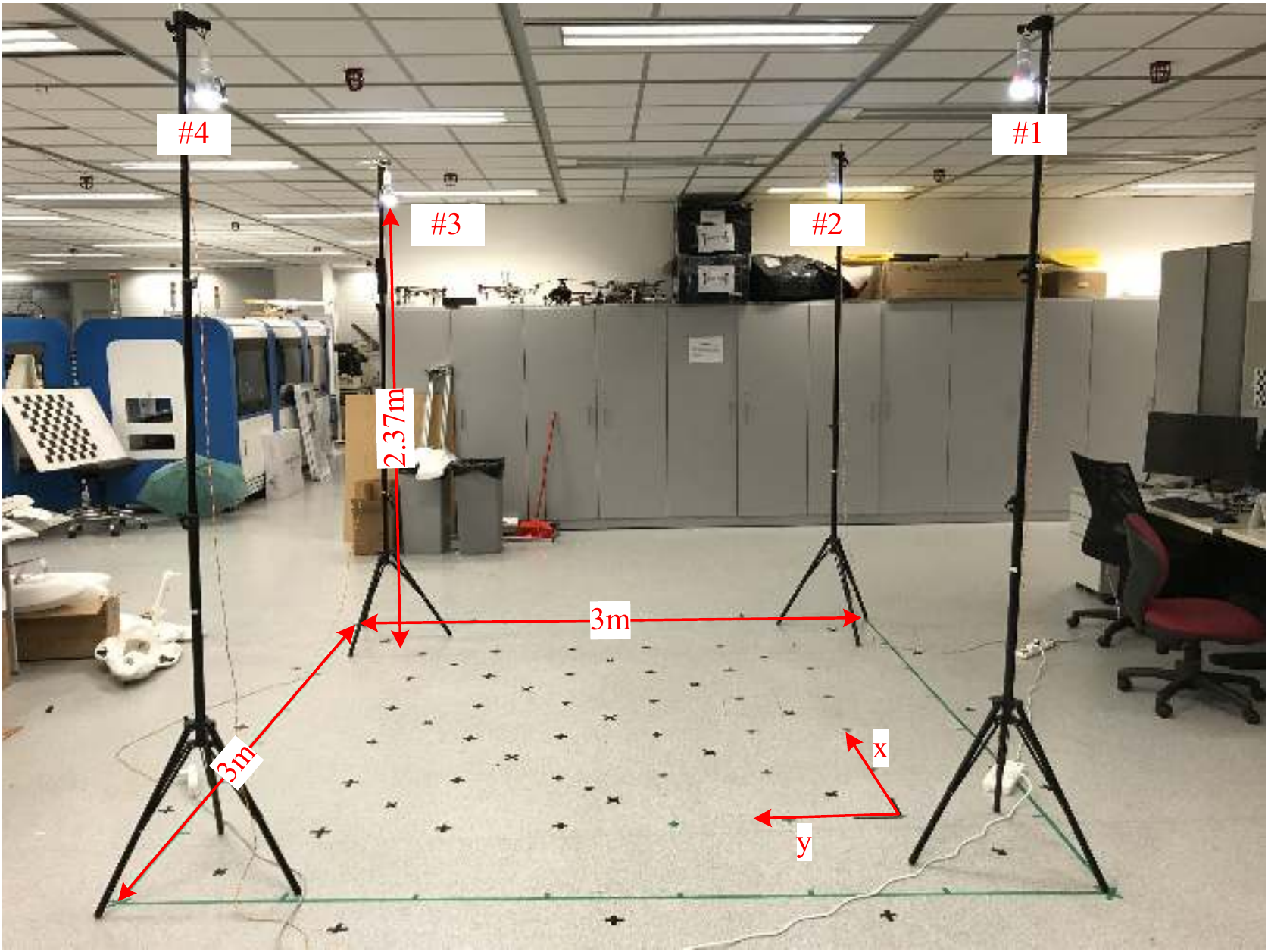}
    \caption{Testbed with four LED bulbs: light\#1, light\#2, light\#3 and light\#4. 	The x-axis and y-axis are labelled with single-end arrows.
    }
    \label{fig:testbed}
\end{figure}     

\begin{figure}[thpb]
	\centering    
    \includegraphics[scale=0.6]{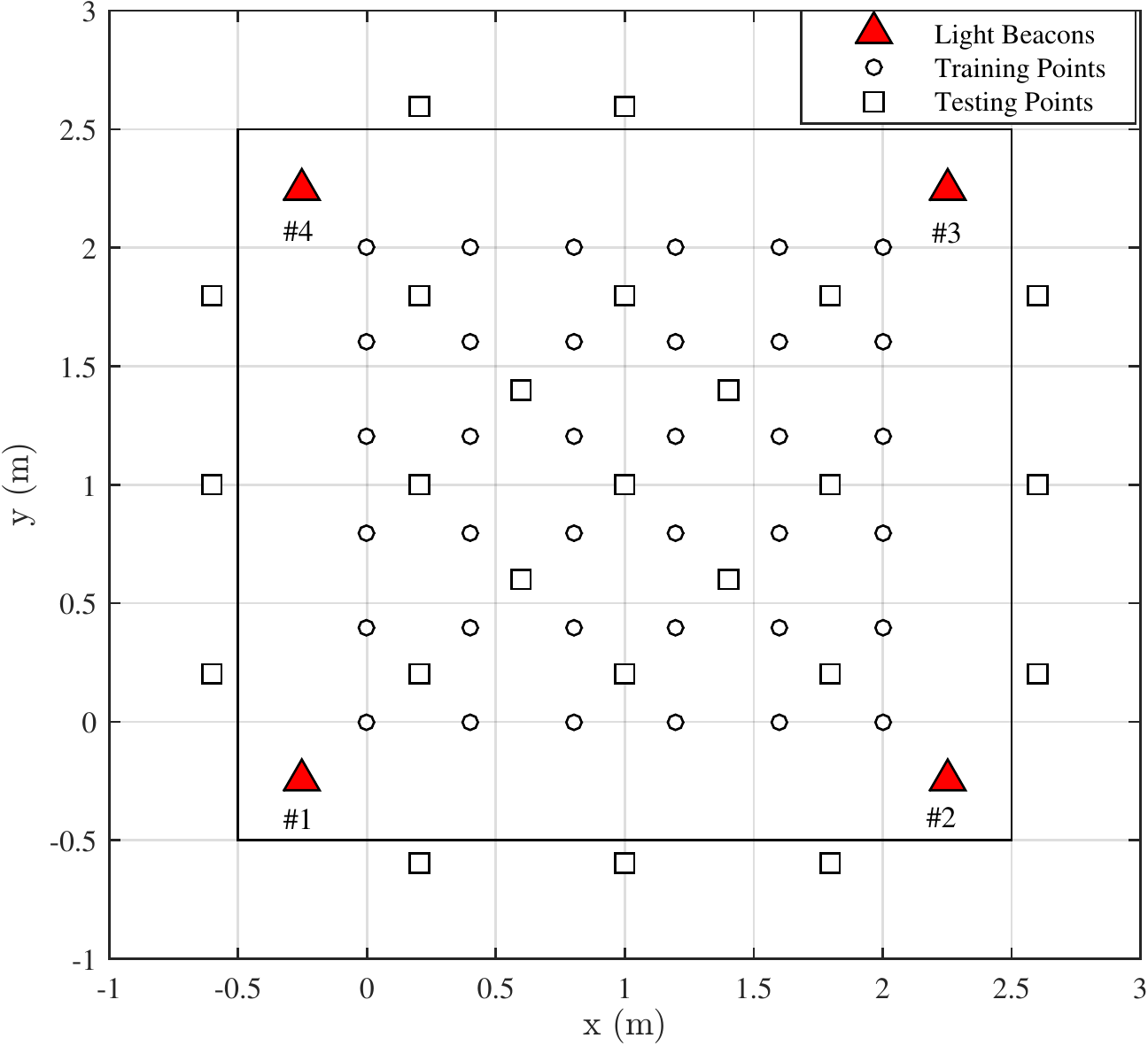}  
    \caption{Testbed floor plan illustrating the training and testing points.}
    \label{fig:floorplan}
\end{figure}

We set up an indoor localization testbed in our lab with four customized LED bulbs, as shown in Figure \ref{fig:testbed}. Four bulbs are installed at the corners of a $3m\times3 m$ square testbed with a height of $2.37m$ over the ground. In the adopted localization algorithm, we use GPR to build a fine-grained light intensity map upon sparse fingerprint samples. We create a 2D grid with $6\times6$ points spaced at $0.4m$ as the training samples for GPR in the central area, as shown in Figure \ref{fig:floorplan}. Then we evenly select 25 extra positions for evaluation covering both the central and border area. We locate these points with an accuracy around $2cm$ with the aid of a commodity laser range finder. 

\subsection{Beacon Broadcasting}
\begin{figure}[thpb]
	\centering
    \includegraphics[scale=0.5]{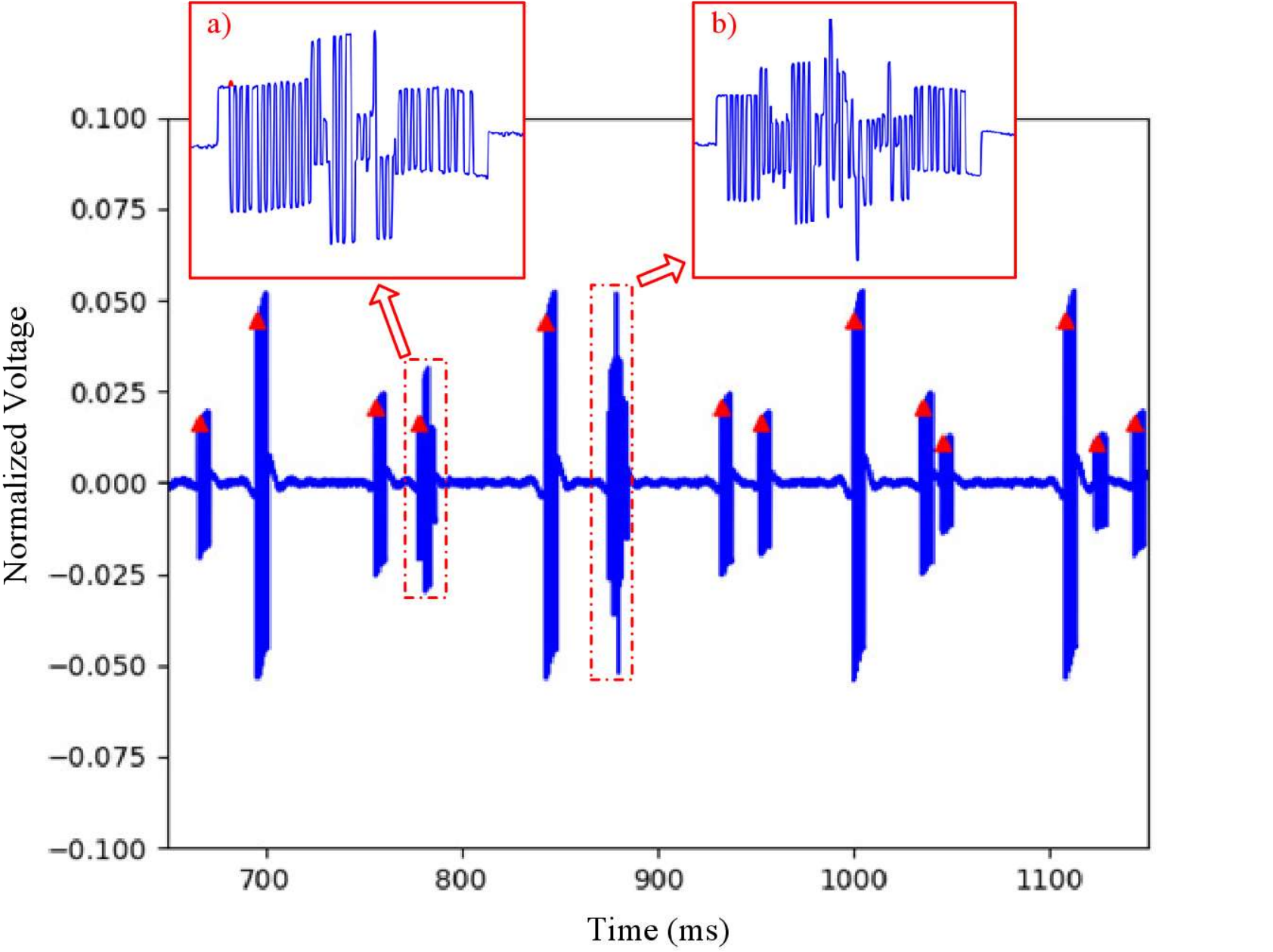}
    \caption{A fragment of raw VLC signals received from four LED bulbs. The decodable messages are marked by red triangles. a) and b) show magnified views of typical corrupted signals. 
    }
    \label{fig:sig_dec}
\end{figure}

Figure \ref{fig:sig_dec} shows a sample of raw VLC signals received simultaneously from four LED bulbs. It is clear that beacon messages are randomly distributed in the time domain. In most cases, they are separated neatly and can be successfully recovered. When collisions occur, as shown in Figure \ref{fig:sig_dec}-b), messages involved will be corrupted. However, we can safely retrieve them by continuous observations later on. Figure \ref{fig:sig_dec}-a) shows a special case. The received message is decodable and checked to be correct. However, there exists a noticeable fluctuation in the signal waveform which is indeed corrupted by others. The RSS measurement is distorted in this situation and may further degrade the localization performance. As a result, we would prefer to discard it. 

\begin{figure}[thpb]
	\centering    
    \includegraphics[scale=0.6]{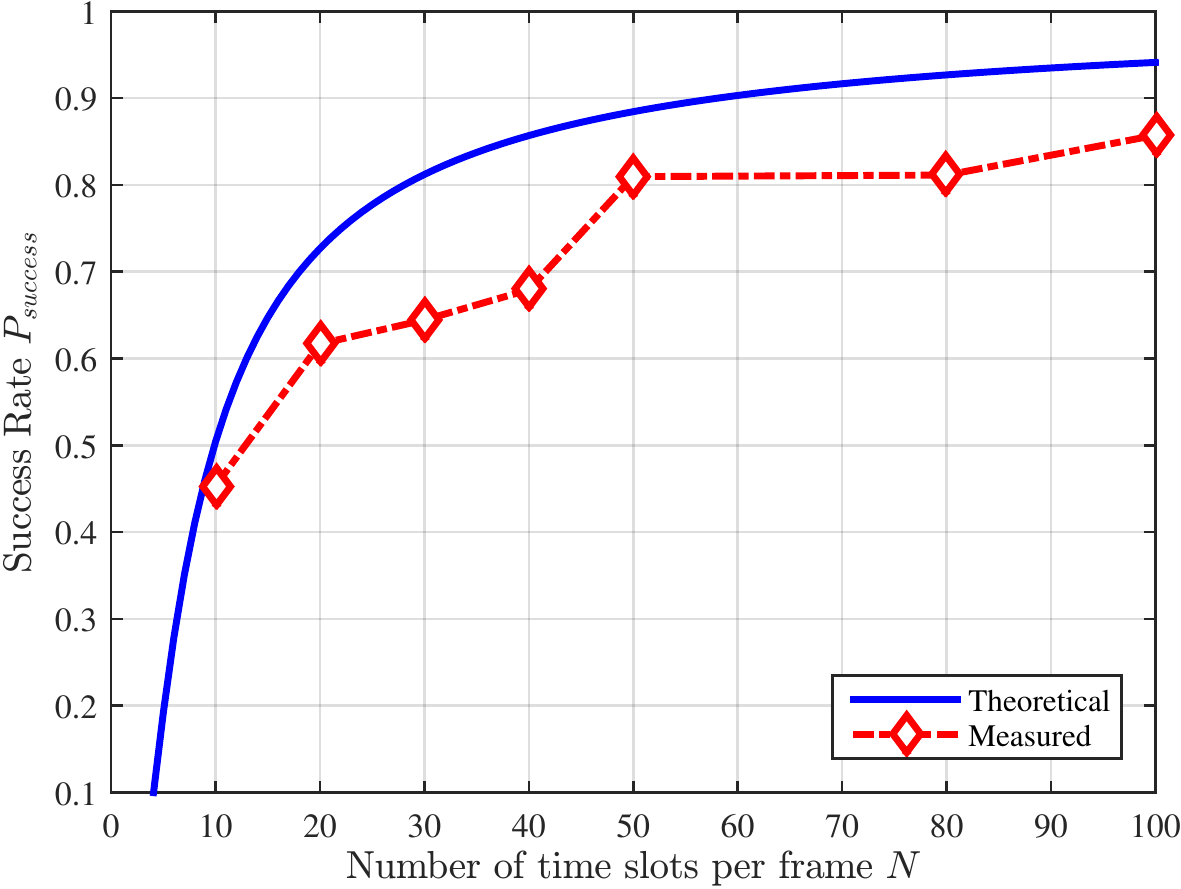}  
    \caption{Success rate versus number of time slots per frame.}
    \label{fig:aloha}
\end{figure} 

We measure the success rate $P_{success}$ of the BFSA-based multiple access scheme and compare it with the theoretical result in Figure \ref{fig:aloha}. The measured values are inferior to the theoretical ones due to the lack of synchronization among the lights. $P_{success}$ increases with the number of time slots per frame $N$. In our implementation, we use a USB soundcard as the ADC on the receiver side with a maximum sampling rate of 48 kHz. To grant reliable signal recovery, the OOK modulation frequency is set to 10 kHz. It takes around 5.6 ms to transmit a full data packet. We empirically choose $N = 20$ so as to provide an acceptable localization latency around $112ms$.

\subsection{Localization with Plugo}
During the experiment, we turn on the four bulbs and keep other lights off. This is because the maximum power (3W) of the LED bulbs is much lower than the normal power rating of other fluorescent lights. The low-frequency components with high energy from these lights will cause the saturation of the receiver circuitry. However, we claim that this will not be a problem if we use higher-power LED lights as we can choose a smaller amplifying gain to prevent the saturation.

\begin{figure}[thpb]
	\centering    
    \includegraphics[scale=0.6]{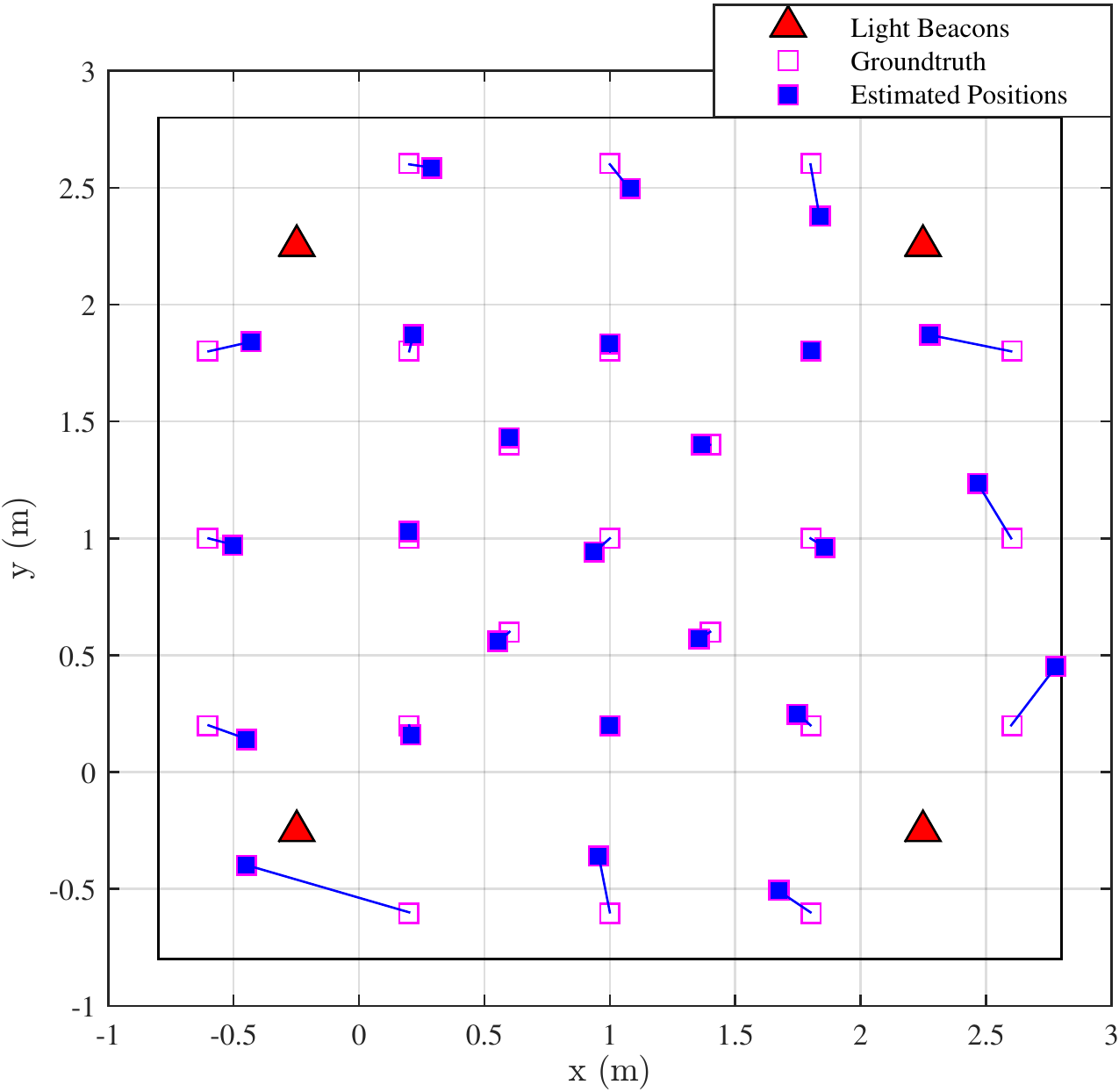}  
    \caption{Localization results at 25 static positions.}
    \label{fig:static}
\end{figure}

\begin{figure}[thpb]
	\centering      
    \includegraphics[scale=0.6]{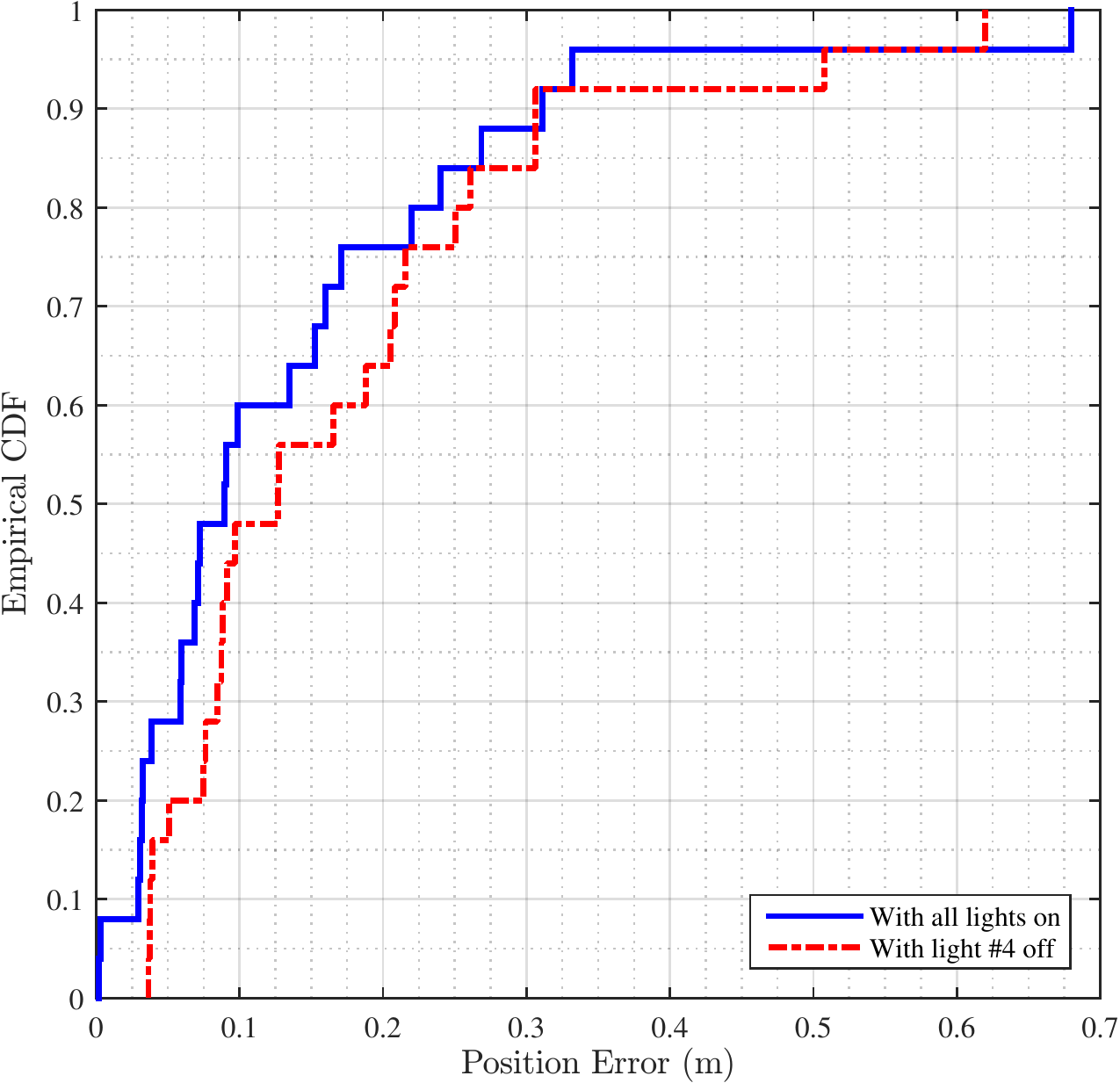}  
    \caption{Position estimation error CDF when four lights are turned on (with solid lines) and one light off (with dashed lines).}
    \label{fig:cdf}
\end{figure}

We collect fingerprint samples at 36 positions and build the intensity distribution maps for all the light beacons. We first conduct the experiment at 25 static positions which are illustrated in Figure \ref{fig:floorplan}. The estimated positions along with the groundtruth are plotted in Figure \ref{fig:static}. The maximum localization errors occur near the testbed borders. This is because we only collect fingerprint samples in the central area of the testbed. The generated light intensity map does not fit well the intensity distribution in the border area. Figure \ref{fig:cdf} plots the empirical CDF of the position errors with the solid curve. The average error is $0.14m$ and the 90-percentile error is $0.33m$. To evaluate the robustness to lights failure, we deliberately switch off light \#4 and redo the experiment. The position error CDF is shown by the dashed line in Figure \ref{fig:cdf}. The localization accuracy is slightly degraded. But we still achieve an average error of $0.17m$ and a 90-percentile error of $0.50m$. It shows that the localization system built upon Plugo is robust to the absence of lights to some extent.

\begin{figure}[thpb]
	\centering	
	\includegraphics[scale=0.6]{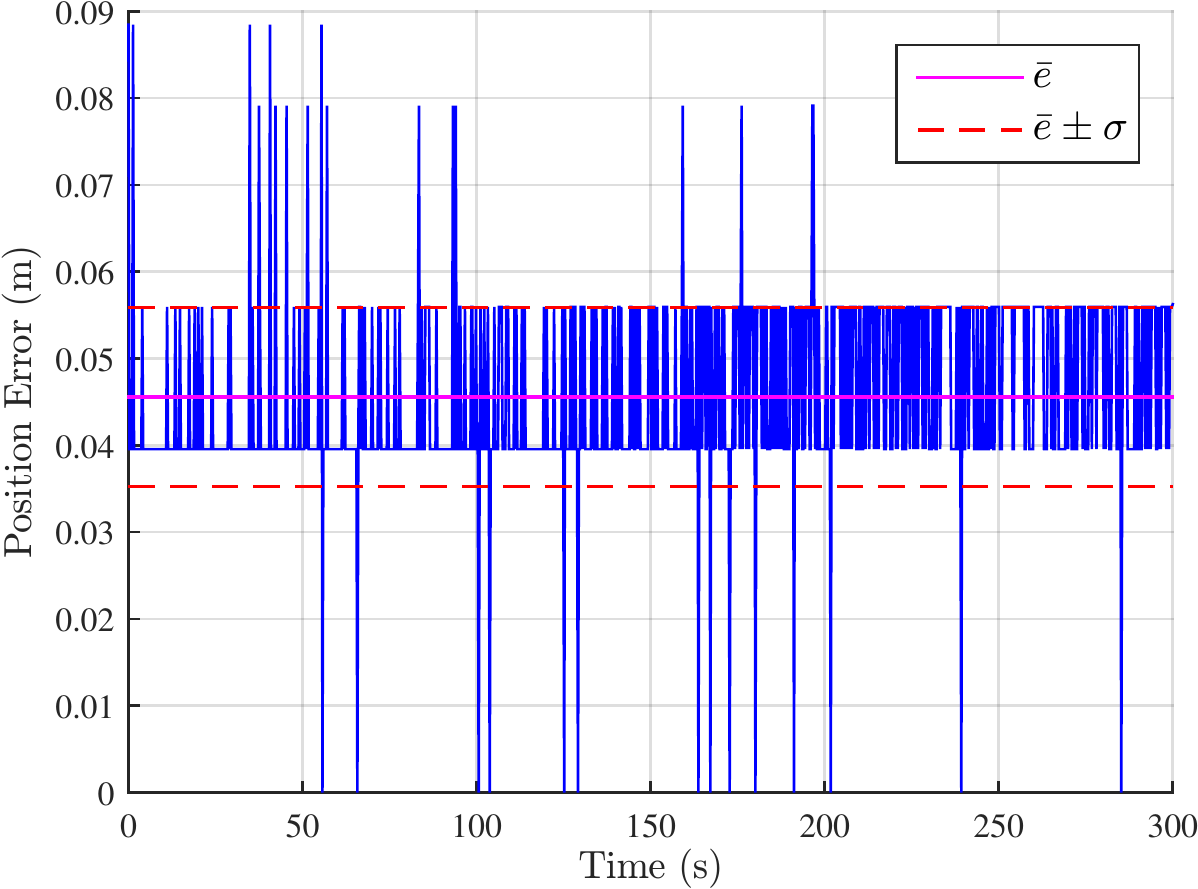}  
    \caption{Localization error at $(1.0, 1.0)$ with respect to time.}
    \label{fig:single}
\end{figure}

\begin{figure}[thpb]
	\centering    
    \includegraphics[scale=0.6]{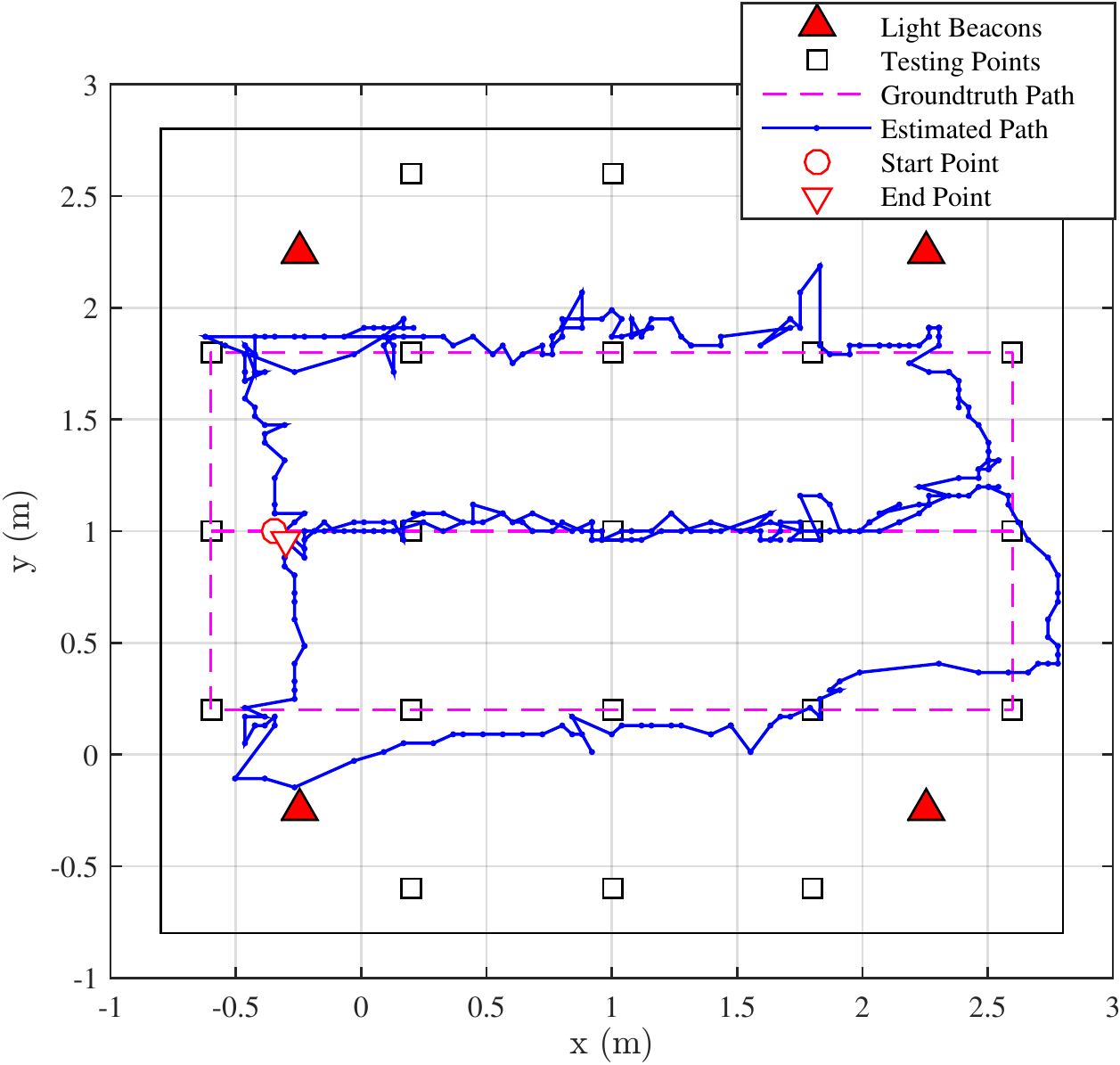}  
    \caption{Continuous localization results along a fixed trajectory.}
    \label{fig:motion}
\end{figure}

We also evaluate the consistency of the localization results. Specifically, we fix the VLC receiver on the center of the testbed and record the estimated positions continuously for 300s, as shown in Figure \ref{fig:single}. The average error is $0.046m$ and the standard deviation is $0.01m$. The error plot appears to be discrete. This is because of the discretization of the intensity distribution maps with a resolution of $0.04m$. It is reasonable that the average error is comparable to the map resolution. The error variation at the static position is relatively small which demonstrates sound system consistency. To demonstrate a real-time use case, we move the receiver on the ground along a fixed trajectory comprising two closed rectangles, as shown in Figure \ref{fig:motion}. The estimated trajectory includes two close-loops similar to the groundtruth. We notice large errors around $0.4m$ near the testbed borders and small errors in the central area.

\subsection{Simulation of Large-scale Scenarios}
Since the hardware-based testbed covers a limited scope of $3m\times3m$ and involves only four LED bulbs, it prevents us from exploring the system scalability to large-scale scenarios. To circumvent this situation, we perform simulations of the beacon broadcasting in a floor-size indoor environment.

\subsubsection{Simulation Setup and Evaluation Metric}
The simulation space is of size $30m \times 30m \times 2.5m$. We place 81 LED bulbs evenly on the ceiling with a separation of 3$m$, and put the receiver on the ground. The vertical distance is fixed to $h=2.5m$ between the receiver and each light. We create a uniform $40\times 40$ grid on the ground and evaluate the beacon broadcasting performance on each grid point. We characterize the radiation pattern of LED bulbs using the first-order Lambertian model\cite{li2014epsilon}, and further assume the photodiode receiver faces upwards and bulbs face downwards. To this end, the channel gains at different evaluation locations could be easily calculated for each LED bulb.

As for trilateration methods, the minimum number of beacons required is three for 2D localization and four for 3D cases. From the state estimation perspective, however, we believe that this requirement can be relaxed. That is, any number of effective measurements of the surrounding environment could probably help improve the location estimation. Considering that lights beyond the FOV or far away from the receiver contribute little to the improvement of localization performance, we empirically select the top four strongest signals for evaluation. They normally come from nearer LED lights at smaller incidence angles so as to induce fewer uncertainties in RSS measurement. To this end, we slightly modify the definition of success rate $P_{success}$ as the total number of successfully received messages over the number of sent messages from the top four strongest signal sources. 

We generate an artificial VLC signal sequence for each LED bulb embedding its unique identification code using OOK modulation with Manchester coding along with the BFSA-based multiple access control. The data frame structure is the same as that used in a real LED bulb. To be specific, we choose $N=20$ time slots per transmission and conduct 20 times of beacon transmissions. The received VLC signal is a linear combination of VLC signals from the lights within the photodiode FOV weighted by the corresponding DC channel gains. Finally, we feed the synthesized signals to the same decoding mechanism as that of field experiments and calculate the success rates. 

\subsubsection{Simulation Results}
\begin{figure}[thpb]
    \centering
    \begin{subfigure}[b]{0.485\columnwidth}
        \includegraphics[width=\textwidth]{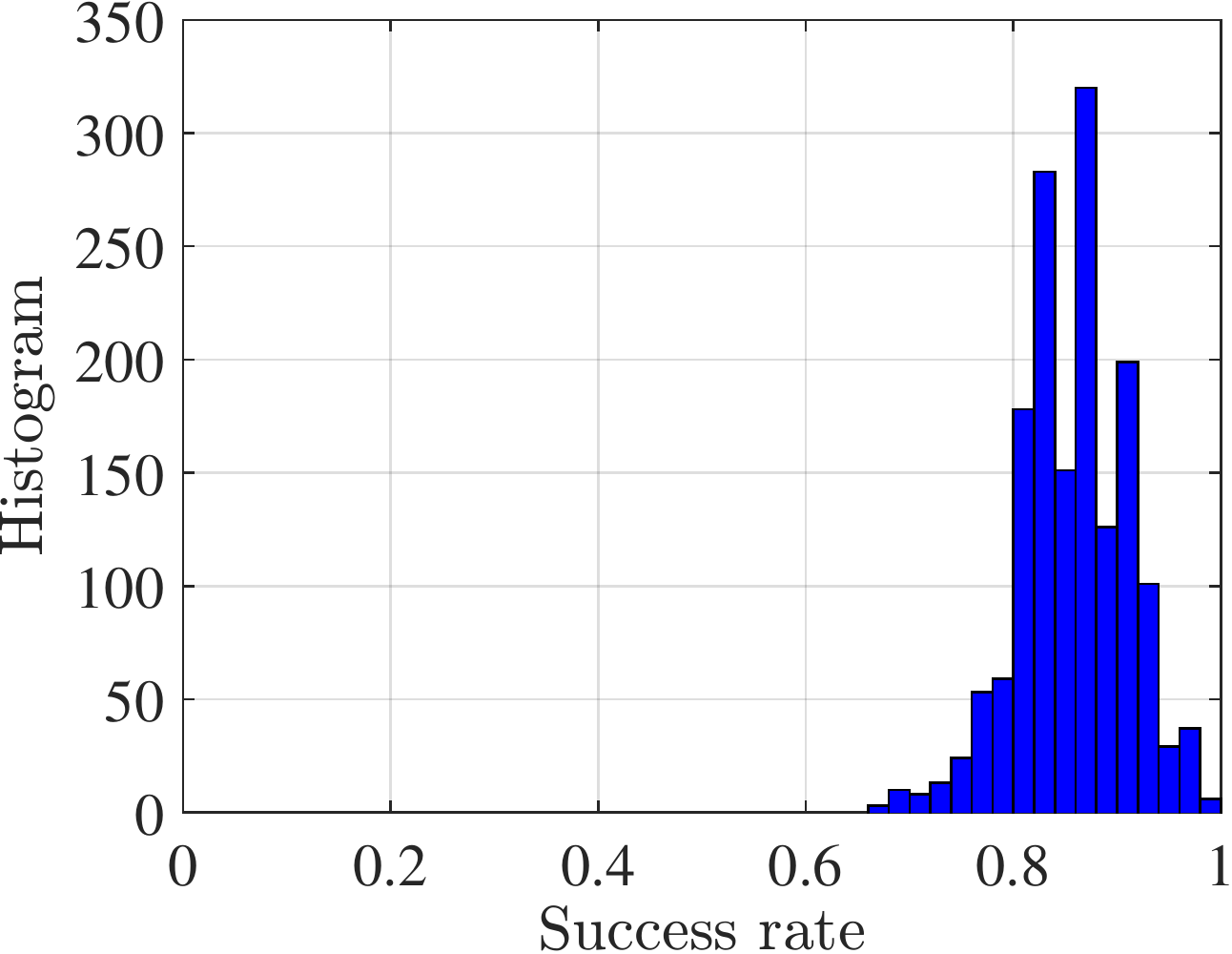}
        \caption{with N=20}
        \label{fig:his20}
    \end{subfigure}
    ~ 
    \begin{subfigure}[b]{0.475\columnwidth}
        \includegraphics[width=\textwidth]{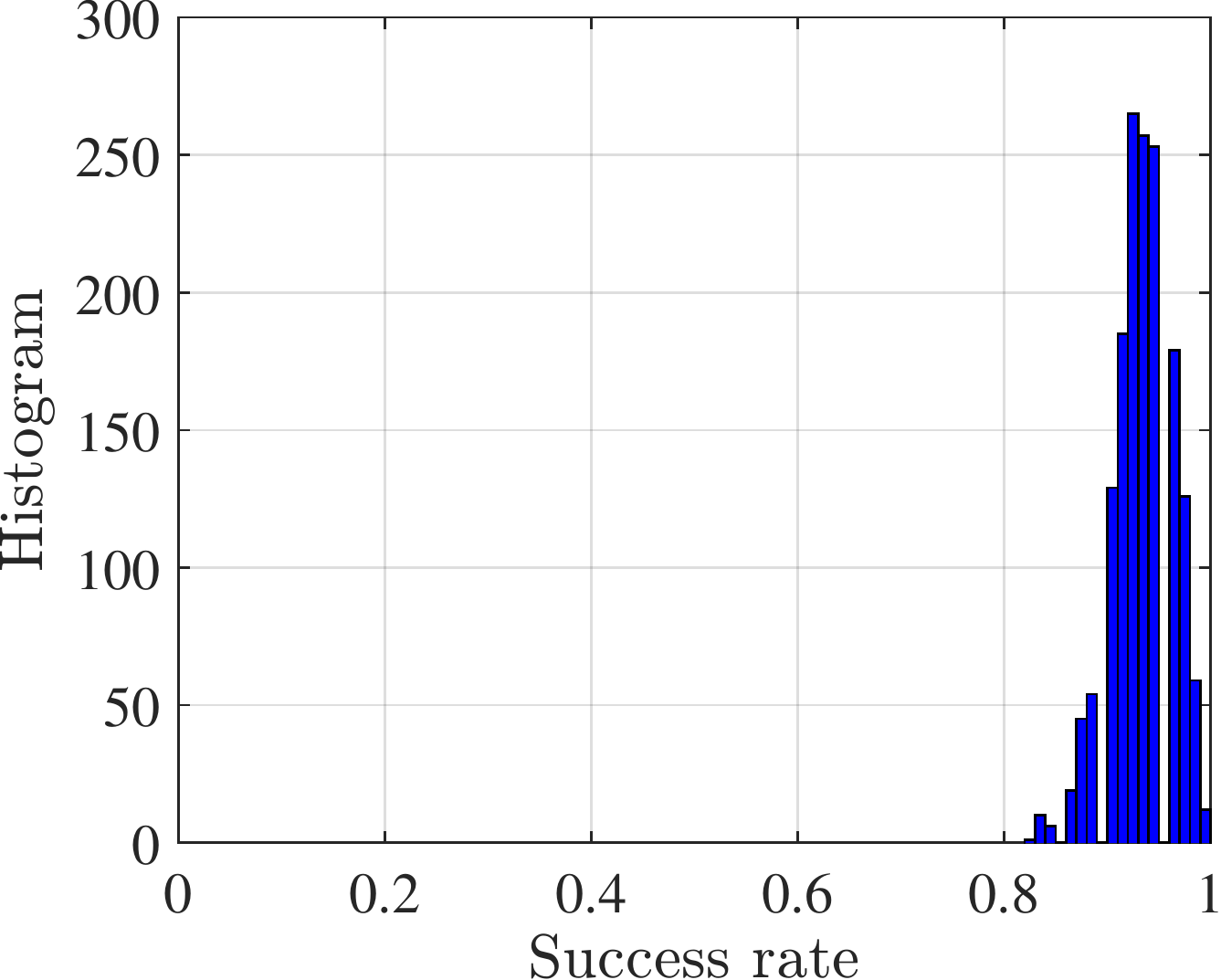}
        \caption{with N=50}
        \label{fig:his50}
    \end{subfigure}
    \caption{Histogram of the success rates at 1600 positions}\label{fig:hist}
\end{figure}

Figure \ref{fig:his20} plots the histogram of the success rates evaluated at 1600 positions with $N=20$. Most success rates center on the median value of $0.85$. The ``tall and thin" shape of the histogram implies reliable beacon broadcasting performance in the floor-size simulation environment. As a result, we may safely consider that the beacon broadcasting performance of Plugo is well scalable to large-scale scenarios. We notice that the average success rate is higher than that measured in the field experiment which also adopts $N=20$ time slots per transmission. This is because the time slots for different LED bulbs is perfectly aligned in the simulation cases. Collisions are thus avoided due to the partial overlap of time slots that are frequently encountered in real applications. In addition, we simulate the noise in beacon broadcasting using an additive white Gaussian noise with a small variance. In the real scenario, however, the noises may be much severe thus degrading the success rate. Then we set $N=50$ and redo the simulation with the results shown in Figure \ref{fig:his50}. The histogram is shifted to the right with a higher median value around $0.93$ along with a much smaller spread. This reveals that the beacon broadcasting performance improves with the number of time slots assigned to each transmission.

\subsection{Discussion}
According to the key technical criteria identified in Section II.B for the expected large-scale indoor localization technologies, we briefly discuss the system performance of Plugo in beacon broadcasting and localization.

\textbf{Accuracy \& Precision:} Through the fingerprinting-based localization algorithm, we demonstrated the feasibility of accurate localization using Plugo. We have achieved an average accuracy of $0.14m$ and a 90-percentile accuracy of $0.33m$. In the literature, the 90-percentile accuracy is normally adopted as the precision evaluation metric. 

\textbf{Responsiveness:} We demonstrated a real-time application of Plugo by tracking a moving receiver along close-loops on the ground with an update rate around 9 Hz. The estimated trajectory showed sound geometry consistency.

\textbf{Scalability:} In Section II, we analyzed qualitatively the system scalability. Moreover, we demonstrated the system scalability in beacon broadcasting to large-scale scenarios by simulation. 

\textbf{Robustness:} We considered a special case of light failure by switching off one of the four lights. The localization accuracy is slightly degraded but still acceptable. The result has revealed, to some extent, the system robustness.

\textbf{Other Criteria:} The remainder of the aforementioned criteria are characterized by ``lightweight", ``low-cost", and ``ubiquitous". We emphasize that, according to previous discussions, they are inherently fulfilled by VLC-based localization systems using photodiode receivers. 

\section{CONCLUSIONS}
In this paper, we presented the design, implementation, and evaluation of Plugo, a dedicated VLC system towards large-scale localization. It was built upon a number of customized commodity LED bulbs and a photodiode receiver. Specially, the bulbs are with compact design favoring the plug-and-go deployment. We conducted an in-depth discussion of the design constraints along with considerations for a photodiode-based VLC system towards large-scale localization. Accordingly, we identified three underlying enablers: 1) distributed architecture, 2) one-way communication, and 3) random multiple access. A BFSA-based random multiple access scheme was implemented with practical issues taken into account. Experiment results showed that Plugo was able to achieve reliable beacon broadcasting over the shared optical medium. In addition, we demonstrated its scalability to large-scale scenarios through simulation. Finally, a preliminary localization result is demonstrated using Plugo in a $3m\times3m$ square testbed showing an average accuracy of $0.14m$ and a 90-percentile accuracy of $0.33m$, which has been greatly improved comparing with the state-of-the-art.


\bibliographystyle{IEEEtran}
\bibliography{references}

\begin{thebibliography}{10}
\providecommand{\url}[1]{#1}
\csname url@samestyle\endcsname
\providecommand{\newblock}{\relax}
\providecommand{\bibinfo}[2]{#2}
\providecommand{\BIBentrySTDinterwordspacing}{\spaceskip=0pt\relax}
\providecommand{\BIBentryALTinterwordstretchfactor}{4}
\providecommand{\BIBentryALTinterwordspacing}{\spaceskip=\fontdimen2\font plus
\BIBentryALTinterwordstretchfactor\fontdimen3\font minus
  \fontdimen4\font\relax}
\providecommand{\BIBforeignlanguage}[2]{{%
\expandafter\ifx\csname l@#1\endcsname\relax
\typeout{** WARNING: IEEEtran.bst: No hyphenation pattern has been}%
\typeout{** loaded for the language `#1'. Using the pattern for}%
\typeout{** the default language instead.}%
\else
\language=\csname l@#1\endcsname
\fi
#2}}
\providecommand{\BIBdecl}{\relax}
\BIBdecl

\bibitem{kuo2014luxapose}
Y.-S. Kuo, P.~Pannuto, K.-J. Hsiao, and P.~Dutta, ``Luxapose: Indoor
  positioning with mobile phones and visible light,'' in \emph{Proceedings of
  the 20th annual international conference on Mobile computing and
  networking}.\hskip 1em plus 0.5em minus 0.4em\relax ACM, 2014, pp. 447--458.

\bibitem{armstrong2013visible}
J.~Armstrong, Y.~Sekercioglu, and A.~Neild, ``Visible light positioning: a
  roadmap for international standardization,'' \emph{IEEE Communications
  Magazine}, vol.~51, no.~12, pp. 68--73, 2013.

\bibitem{lumicast}
J.~Aleksandar, ``A high accuracy indoor positioning system based on visible
  light communication,''
  \url{https://www.qualcomm.com/media/documents/files/lumicast-whitepaper.pdf},
  accessed June 9, 2017.

\bibitem{marketreport}
MarketsandMarkets, ``Indoor location market by component (technology, software
  tools, and services), application, end user (transportation, hospitality,
  entertainment, shopping, and public buildings), and region - global forecast
  to 2021,''
  \url{http://www.marketsandmarkets.com/Market-Reports/indoor-positioning-navigation-ipin-market-989.html},
  accessed June 9, 2017.

\bibitem{liu2007survey}
H.~Liu, H.~Darabi, P.~Banerjee, and J.~Liu, ``Survey of wireless indoor
  positioning techniques and systems,'' \emph{IEEE Transactions on Systems,
  Man, and Cybernetics, Part C (Applications and Reviews)}, vol.~37, no.~6, pp.
  1067--1080, 2007.

\bibitem{huang2011efficient}
J.~Huang, D.~Millman, M.~Quigley, D.~Stavens, S.~Thrun, and A.~Aggarwal,
  ``Efficient, generalized indoor wifi graphslam,'' in \emph{Robotics and
  Automation (ICRA), 2011 IEEE International Conference on}.\hskip 1em plus
  0.5em minus 0.4em\relax IEEE, 2011, pp. 1038--1043.

\bibitem{sun2014wifi}
Y.~Sun, M.~Liu, and M.~Q.-H. Meng, ``Wifi signal strength-based robot indoor
  localization,'' in \emph{Information and Automation (ICIA), 2014 IEEE
  International Conference on}.\hskip 1em plus 0.5em minus 0.4em\relax IEEE,
  2014, pp. 250--256.

\bibitem{Cadena2016}
\BIBentryALTinterwordspacing
C.~Cadena, L.~Carlone, H.~Carrillo, Y.~Latif, D.~Scaramuzza, J.~Neira, I.~Reid,
  and J.~Leonard, ``{Past, Present, and Future of Simultaneous Localization and
  Mapping: Toward the Robust-Perception Age},'' \emph{IEEE Transactions on
  Robotics}, vol.~32, no.~6, pp. 1309--1332, 2016. [Online]. Available:
  \url{http://ieeexplore.ieee.org/document/7747236/}
\BIBentrySTDinterwordspacing

\bibitem{zhong2016efficient}
C.~Zhong, S.~Liu, Q.~Lu, B.~Zhang, and S.~X. Yang, ``An efficient
  fine-to-coarse wayfinding strategy for robot navigation in regionalized
  environments,'' \emph{IEEE transactions on cybernetics}, vol.~46, no.~12, pp.
  3157--3170, 2016.

\bibitem{yang2004neural}
S.~X. Yang and C.~Luo, ``A neural network approach to complete coverage path
  planning,'' \emph{IEEE Transactions on Systems, Man, and Cybernetics, Part B
  (Cybernetics)}, vol.~34, no.~1, pp. 718--724, 2004.

\bibitem{do2016depth}
T.-H. Do and M.~Yoo, ``An in-depth survey of visible light communication based
  positioning systems,'' \emph{Sensors}, vol.~16, no.~5, p. 678, 2016.

\bibitem{yang2015wearables}
Z.~Yang, Z.~Wang, J.~Zhang, C.~Huang, and Q.~Zhang, ``Wearables can afford:
  Light-weight indoor positioning with visible light,'' in \emph{Proceedings of
  the 13th Annual International Conference on Mobile Systems, Applications, and
  Services}.\hskip 1em plus 0.5em minus 0.4em\relax ACM, 2015, pp. 317--330.

\bibitem{li2014epsilon}
L.~Li, P.~Hu, C.~Peng, G.~Shen, and F.~Zhao, ``Epsilon: A visible light based
  positioning system.'' in \emph{NSDI}, 2014, pp. 331--343.

\bibitem{zhang2014asynchronous}
W.~Zhang, M.~S. Chowdhury, and M.~Kavehrad, ``Asynchronous indoor positioning
  system based on visible light communications,'' \emph{Optical Engineering},
  vol.~53, no.~4, pp. 045\,105--045\,105, 2014.

\bibitem{qiu2016let}
K.~Qiu, F.~Zhang, and M.~Liu, ``Let the light guide us: Vlc-based
  localization,'' \emph{IEEE Robotics \& Automation Magazine}, vol.~23, no.~4,
  pp. 174--183, 2016.

\bibitem{rajagopal2012ieee}
S.~Rajagopal, R.~D. Roberts, and S.-K. Lim, ``Ieee 802.15. 7 visible light
  communication: modulation schemes and dimming support,'' \emph{IEEE
  Communications Magazine}, vol.~50, no.~3, 2012.

\bibitem{zhang2015asynchronous}
F.~Zhang, K.~Qiu, and M.~Liu, ``Asynchronous blind signal decomposition using
  tiny-length code for visible light communication-based indoor localization,''
  in \emph{Robotics and Automation (ICRA), 2015 IEEE International Conference
  on}.\hskip 1em plus 0.5em minus 0.4em\relax IEEE, 2015, pp. 2800--2805.

\bibitem{chen2015framework}
Y.-A. Chen, Y.-T. Chang, Y.-C. Tseng, and W.-T. Chen, ``A framework for
  simultaneous message broadcasting using cdma-based visible light
  communications,'' \emph{IEEE Sensors Journal}, vol.~15, no.~12, pp.
  6819--6827, 2015.

\bibitem{karunatilaka2015led}
D.~Karunatilaka, F.~Zafar, V.~Kalavally, and R.~Parthiban, ``Led based indoor
  visible light communications: State of the art.'' \emph{IEEE Communications
  Surveys and Tutorials}, vol.~17, no.~3, pp. 1649--1678, 2015.

\bibitem{qiu2015visible}
K.~Qiu, F.~Zhang, and M.~Liu, ``Visible light communication-based indoor
  localization using gaussian process,'' in \emph{Intelligent Robots and
  Systems (IROS), 2015 IEEE/RSJ International Conference on}.\hskip 1em plus
  0.5em minus 0.4em\relax IEEE, 2015, pp. 3125--3130.

\bibitem{yasir2014indoor}
M.~Yasir, S.-W. Ho, and B.~N. Vellambi, ``Indoor positioning system using
  visible light and accelerometer,'' \emph{Journal of Lightwave Technology},
  vol.~32, no.~19, pp. 3306--3316, 2014.

\bibitem{wang2015open}
Q.~Wang, D.~Giustiniano, and D.~Puccinelli, ``An open source research platform
  for embedded visible light networking,'' \emph{IEEE Wireless Communications},
  vol.~22, no.~2, pp. 94--100, 2015.

\bibitem{klaver2015shine}
L.~Klaver and M.~Zuniga, ``Shine: A step towards distributed multi-hop visible
  light communication,'' in \emph{Mobile Ad Hoc and Sensor Systems (MASS), 2015
  IEEE 12th International Conference on}.\hskip 1em plus 0.5em minus
  0.4em\relax IEEE, 2015, pp. 235--243.

\bibitem{philips}
Philips, ``Unlocking the value of retail apps with lighting,''
  \url{http://www.lighting.philips.com/main/systems/themes/led-based-indoor-positioning/white-paper},
  accessed June 9, 2017.

\bibitem{kim2013indoor}
H.-S. Kim, D.-R. Kim, S.-H. Yang, Y.-H. Son, and S.-K. Han, ``An indoor visible
  light communication positioning system using a rf carrier allocation
  technique,'' \emph{Journal of Lightwave Technology}, vol.~31, no.~1, pp.
  134--144, 2013.

\bibitem{liu2014towards}
M.~Liu, K.~Qiu, F.~Che, S.~Li, B.~Hussain, L.~Wu, and C.~P. Yue, ``Towards
  indoor localization using visible light communication for consumer electronic
  devices,'' in \emph{Intelligent Robots and Systems (IROS 2014), 2014 IEEE/RSJ
  International Conference on}.\hskip 1em plus 0.5em minus 0.4em\relax IEEE,
  2014, pp. 143--148.

\bibitem{jplrandom}
L.~Jean, Paul, ``Jpl's wireless communication reference website,''
  \url{http://www.wirelesscommunication.nl/reference/chaptr06/randacc.htm},
  accessed June 9, 2017.

\bibitem{ieeevlc}
``Ieee standard for local and metropolitan area networks--part 15.7:
  Short-range wireless optical communication using visible light,'' \emph{IEEE
  Std 802.15.7-2011}, pp. 1--309, Sept 2011.

\bibitem{rasmussen2006gaussian}
C.~E. Rasmussen, ``Gaussian processes for machine learning,'' 2006.

\end{thebibliography}

%
\begin{IEEEbiography}[{\includegraphics[width=1in,height=1.25in,clip,keepaspectratio]{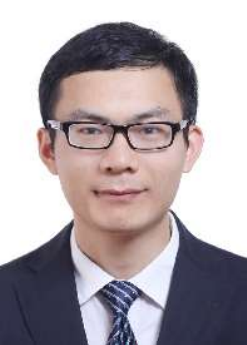}}]{Qing Liang}
received the B.A. degree in Automation at Xi'an Jiaotong University (XJTU), Xi'an, China, in 2013 and the master degree in Instrument Science and Technology at Beihang University (BUAA), Beijing, China, in 2016. He is now a Ph.D. student at the Department of Mechanical and Biomedical Engineering, City University of Hong Kong. His current research interests include sensor fusion, low-cost localization, and mobile robots.
\end{IEEEbiography}

\begin{IEEEbiography}[{\includegraphics[width=1in,height=1.25in,clip,keepaspectratio]{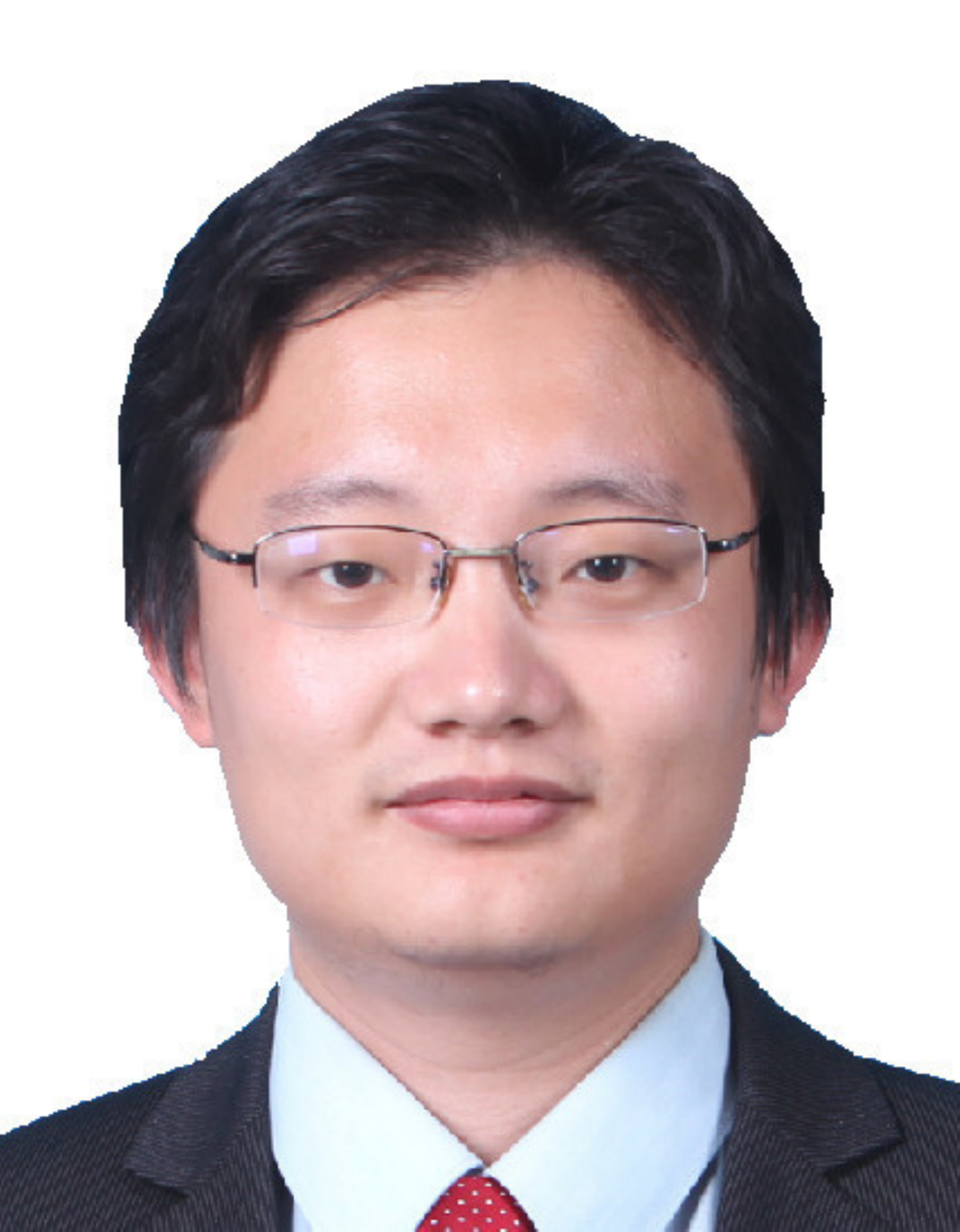}}]{Ming Liu}
received the B.A. degree in Automation at Tongji University in 2005. During his master study at Tongji University, he stayed one year in Erlangen-Nünberg University and Fraunhofer Institute IISB, Germany, as a master visiting scholar. He graduated as a PhD student from the Department of Mechanical and Process Engineering of ETH Zürich in 2013. 

He is now affiliated with ECE department, CSE department and Robotics Institute of Hong Kong University of Science and Technology. His research interests include dynamic environment modeling, deep-learning for robotics, 3D mapping, machine learning and visual control. 

Prof. Liu is the recipient of the Best Student Paper Award at IEEE MFI 2012, the Best Paper in Information Award at IEEE ICIA 2013, the Best RoboCup Paper at IEEE IROS 2013, and twice the Winning Prize of the Chunhui-Cup Innovation Contest.
\end{IEEEbiography}

\end{document}